\documentclass[a4paper,12pt]{article}
\usepackage[utf8]{inputenc}
\usepackage{amsfonts}
\usepackage{amssymb}
\usepackage{graphicx}
\usepackage{amsmath}
\usepackage[table,xcdraw]{xcolor}
\usepackage{tikz}
\usetikzlibrary{decorations.markings}
\usepackage{enumerate}

\usepackage{cases}
\usepackage{subfig}
\usepackage{pgfplots}
\usepackage{color}
\usepackage{tikz}\usetikzlibrary{calc}
\usepackage{float}
\usepackage{here}
\usepackage{cite}
\usepackage{mathrsfs}
\usepackage{float,epsfig}
\usepackage{dcolumn}% Align table columns on decimal point

\usepackage{graphicx}% Include figure files
\usepackage{bm}% bold math
\usepackage{amsmath,amssymb,amsthm}
\usepackage[colorlinks=true,linkcolor=blue,citecolor=red]{hyperref}
\newcommand{\f}{\frac}

\newcommand{\rt}{\right}
\newcommand{\be}{\begin{equation}}
\newcommand{\ee}{\end{equation}}
\newcommand{\bea}{\setlength\arraycolsep{2pt} \begin{eqnarray}}
\newcommand{\eea}{\end{eqnarray}}

\def\0{{\sst{(0)}}}
\def\1{{\sst{(1)}}}
\def\2{{\sst{(2)}}}
\def\3{{\sst{(3)}}}
\def\4{{\sst{(4)}}}
\def\5{{\sst{(5)}}}
\def\6{{\sst{(6)}}}
\def\7{{\sst{(7)}}}
\def\8{{\sst{(8)}}}
\def\sst#1{{\scriptscriptstyle #1}}

\newcommand{\pgftextcircled}[1]{
    \setbox0=\hbox{#1}%
    \dimen0\wd0%
    \divide\dimen0 by 2%
    \begin{tikzpicture}[baseline=(a.base)]%
        \useasboundingbox (-\the\dimen0,0pt) rectangle (\the\dimen0,1pt);
        \node[circle,draw,outer sep=0pt,inner sep=0.1ex] (a) {#1};
    \end{tikzpicture}
}

\definecolor{lime}{HTML}{A6CE39}
\newcommand{\orcidicon}{%
    \begin{tikzpicture}
    \draw[lime, fill=lime] (0,0)
        circle [radius=0.16]
        node[white] {{\fontfamily{qag}\selectfont \tiny ID}};
    \draw[white, fill=white] (-0.0625,0.095)
        circle [radius=0.007];
    \end{tikzpicture}   \hspace{-2mm}
}
\newcommand\orcidAdil{{\href{https://orcid.org/0000-0001-7623-5541}{\orcidicon}}}

\newcommand\orcidMohamed{{\href{https://orcid.org/0000-0003-1185-0062}{\orcidicon}}}
\newcommand\orcidHajar{{\href{https://orcid.org/0000-0001-9510-4248}{\orcidicon}}}

\makeatletter \@addtoreset{equation}{section}

\usepackage{multirow}
\newcommand*\frontaleye{%
       \scalebox{0.25}{
\tikzset{every picture/.style={line width=0.75pt}}
\begin{tikzpicture}[x=0.75pt,y=0.75pt,yscale=-1,xscale=1]
\draw  [draw opacity=0][fill={rgb, 255:red, 140; green, 196; blue, 74 }  ,fill opacity=1 ] (104.5,179.99) .. controls (104.5,171.39) and (111.48,164.41) .. (120.08,164.41) .. controls (128.68,164.41) and (135.66,171.39) .. (135.66,179.99) .. controls (135.66,188.6) and (128.68,195.57) .. (120.08,195.57) .. controls (111.48,195.57) and (104.5,188.6) .. (104.5,179.99) -- cycle ;
\draw  [fill={rgb, 255:red, 0; green, 0; blue, 0 }  ,fill opacity=1 ] (84.83,179.7) .. controls (84.92,169.21) and (100.66,160.85) .. (119.99,161.01) .. controls (139.32,161.18) and (154.92,169.81) .. (154.83,180.3) .. controls (152.49,171.04) and (137.8,163.81) .. (119.97,163.66) .. controls (102.13,163.51) and (87.32,170.48) .. (84.83,179.7) -- cycle ;
\draw  [fill={rgb, 255:red, 0; green, 0; blue, 0 }  ,fill opacity=1 ] (113.6,179.91) .. controls (113.6,176.19) and (116.63,173.16) .. (120.36,173.16) .. controls (124.08,173.16) and (127.11,176.19) .. (127.11,179.91) .. controls (127.11,183.64) and (124.08,186.67) .. (120.36,186.67) .. controls (116.63,186.67) and (113.6,183.64) .. (113.6,179.91) -- cycle ;
\draw  [fill={rgb, 255:red, 0; green, 0; blue, 0 }  ,fill opacity=1 ] (154.65,179.7) .. controls (154.65,190.58) and (139.02,199.41) .. (119.74,199.41) .. controls (100.46,199.41) and (84.83,190.58) .. (84.83,179.7) .. controls (87.23,189.3) and (101.95,196.67) .. (119.74,196.67) .. controls (137.53,196.67) and (152.25,189.3) .. (154.65,179.7) -- cycle ;
\draw  [draw opacity=0][fill={rgb, 255:red, 255; green, 255; blue, 255 }  ,fill opacity=1 ] (125.45,172.24) .. controls (125.45,170.4) and (126.94,168.91) .. (128.78,168.91) .. controls (130.62,168.91) and (132.11,170.4) .. (132.11,172.24) .. controls (132.11,174.08) and (130.62,175.57) .. (128.78,175.57) .. controls (126.94,175.57) and (125.45,174.08) .. (125.45,172.24) -- cycle ;
\end{tikzpicture}
}\kern-.5em}

\begin{document}
%	\begin{flushright}
%	ksdjflk kjsdlfkj
%	\end{flushright}
%

\title{\normalsize
%\phantom{fff}
%\vspace{-3cm}
%\begin{flushright}
%FISPAC-TH/27/2020\\
%UQBAR-TH/2020-314
%\end{flushright}
%\vspace{2cm}
%%%
%%
%%
{\bf \Large	  Light  Trajectories and  Thermal Shadows casted by   Black Holes  in  a Cavity }}
\author{ \small   A. Belhaj\orcidAdil\!\! $^{1}$\footnote{a-belhaj@um5r.ac.ma},  H. Belmahi\orcidHajar\!\!$^{1}$\footnote{hajar\_belmahi@um5.ac.ma},  M. Benali\orcidMohamed\!\!$^{1}$\footnote{mohamed\_benali4@um5.ac.ma}, M. Oualaid$^{1}$\footnote{mohamed\_oualaid@um5.ac.ma}, M. B. Sedra$^{2}$\footnote{mysedra@yahoo.fr}
	\thanks{ Authors in alphabetical order.}
	\hspace*{-8pt} \\
	%EndAName
	{\small $^1$ D\'{e}partement de Physique, Equipe des Sciences de la mati\`ere et du rayonnement, ESMaR}\\
{\small   Facult\'e des Sciences, Universit\'e Mohammed V de Rabat, Rabat,  Morocco} \\
	{\small $^{2}$  Material and subatomic physics laboratory, LPMS,   University of  Ibn Tofail, Kenitra,  Morocco} 
} \maketitle

 \maketitle

\maketitle

%\date{}
%\vspace{-3.6em}
%
%\begin{center}
%\textit{$^1$
%\\ [0.5em]
%\end{center}
%
%\vspace{1em}

	\begin{abstract}

	We explore   the  shadows and the   photon rings casted by black holes  in  cavity. Placing the observer  inside such an   isothermal  background,  we examine  the influence of the  cavity temperature $T_{cav}$ and the charge $Q$ on the  involved optical aspect. After studying the effect of the horizon radius  by varying  $Q$, we investigate  the thermal behaviors of the  black hole  shadows in a cavity.  For fixed   charge values, we find   that the shadow radius $r_s$  increases by decreasing $T_{cav}$.      Varying such a temperture,    we discuss the associated energy emission rate.  After that,  we   show  that the  curves   in  the $ r_s-T_{cav}$ plane   share   similarities with  the $G-T$ curves of  the  Anti de Sitter  (AdS)  black holes.
	Then, we study the trajectory of the light rays  casted by black holes in a  cavity.   We  further observe that  the  light trajectory behaviors  are different  than    the  ones of the non rotating black holes due to  the cavity effect. Finally,  we provide evidence for the existence of  an  universal  ratio defined in terms of  the photon sphere radius and the impact parameter. Concretely, we obtain a optical  ratio   $\frac{b_{sp}}{r_{sp}} \sim \sqrt 3.$

		{\noindent}
	\end{abstract}
\newpage

\tableofcontents
\newpage

\section{Introduction}

Recently,   the study of  the optical and   the thermodynamics of the  black hole has brought many interesting results associated with classical  and quantum gravity models. These findings have been supported by detections and observational investigations of such fascinating objects.  Concretely,  the Even Horizon Telescope (EHT)   international  collaboration  has provided an image of  an  accretion flow through  the vicinity  of  the supermassive black hole  in M87$^*$\cite{a1,a2,a3}. In this image, a dark interior  has been observed. This region called shadow is surrounded by a bright ring known as the photon sphere. Since the elaboration of  such an image and the associated data by EHT, various works dealing with the geometric properties of the black hole shadows have been proposed  by examining the involved sizes and shapes\cite{a4,a5,a6,a7,a8,a9,a10,a11,A11,a12,a13,a14,a140,a141,a142, a143}.  In four dimensions, it has been shown that the non-rotating black hole solutions exhibit a  perfect circular geometry where the size can be controlled by certain parameters  as  the charge\cite{a7,a15,a16}. For  the rotating black holes, however, the circular configuration  is  deformed and distorted generating non trivial  shapes including D and cardioid ones\cite{a17,a18,a19,a20}. It has been remarked that EHT data can be exploited to impose constraints on certain black hole parameters.   These constraints could be used to  built models matching with the EHT observations considered as a reference for testing the obtained theoretical results.

Beside  the optical aspect, the  black hole thermodynamics has received also a remarkable interest  showing non trivial results corresponding to   the criticality and the  stability behaviors.  A crucial focus   has been  devoted to the study of  the black holes on   the AdS geometries\cite{a21,a22,a23,a24}.  The needed quantities have been nicely computed for various gravity theories. Considering the cosmological as a pressure, certain black holes  show similarities with van der Waals  fluid systems where some universalities have been obtained.    It  has been revealed  that   the AdS black holes can
be in  a stable thermal equilibrium with radiations \cite{a21,a23,a26,a27}. Hawking and Page (HP)  have provided theoretical
evidence for the existence of certain  transitions in the phase space of the (non-rotating
uncharged) Schwarzschild-AdS black hole.  A first order phase transition in the
charged (non-rotating) Reissner {Nordstrom-AdS (RN-AdS) black hole space-time has been 
studied in different backgrounds including   Dark Energy (DE) and Dark Matter (DM)\cite{a15,a28,a29}.  The effect of such a  dark sector  on the optical aspect has been also investigated  for different backgrounds.  In this way,  the  shadows and  the deflection angle of the  light rays by black holes in  arbitrary dimensions   have been discussed  in \cite{A11,a15,a16,a17}.   The effect of  DE and the space-time dimension on the involved  optical quantities has been inspected.  In \cite{a18,a20},  for instance, the 
 influence of DM on  the shadows and  the  photon rings of a  stringy black hole illuminated by certain  accretions has been also studied.
 
 Motivated  by  the  investigations on singular space-times,   interplays  between the  black hole thermodynamics and the  optical properties have been established.  It has been shown that the shadow size can provide information on the  HP  phase transitions, the  critical behaviors and  the microstructure states of  the  black holes living in  AdS  geometries \cite{a290,a291,a292}. It can reflect also data  on the geothermodynamics by providing certain universalities \cite{a293}.    It has been shown that the relation between  the shadow and  the thermodynamics of the   black hole has been also developed for regular space-times\cite{a294}.  Using the elliptic function analysis,  it has been explored further  a fundamental connection between the AdS black hole thermodynamics and the deflection angle of the light rays. Concretely,   various  thermodynamics behaviors of such black holes have been approached  in terms   the deflection angle variations \cite{a295}.

More recently,  it has been shown that  the black holes on the   AdS geometries  share  similarities with  certain  black holes enclosed by  an  isothermal cavity. The latter  has been considered as a tool to provide a stable  solution\cite{a30,A30,a31,a32,a33}. Precisely,   it has been  revealed that the  Schwarzschild black holes in a cavity can be thermally stable. They involve  the 
phase structures and transition behaviors similar to the ones appearing in the  Schwarzschild-AdS black holes.   A similar interplay has been observed  in  the Reissner-
Nordstrom (RN) black holes,  by considering  
 canonical ensembles. The phase structures of certain extended models  in a cavity have been  investigated where   the Hawking-Page-like and  van der
Waals-like phase transitions have taken place\cite{a34}. 

The aim of this work is to explore for the first time  the optical behaviors of the black holes in a cavity.  Precisely,  we  investigate the  shadows and the  photon  rings casted by     black holes  surrounded by such an isothermal generic space,  by examining  the effect of the cavity  temperature $T_{cav}$ and the charge $Q$.   After studying the effect of the horizon radius on  the  visualizing  shadows of the proposed 
black holes by varying the charge, we move to  study  thermal behaviors of  the involved  shadows.  For fixed charge values, we observe   that the shadow radius $r_s$  increases by decreasing  the cavity temperature $T_{cav}$.   Varying such a temperature,     we discuss  the energy emission rate.  Moreover,  we  show  that the  curves   in the  $r_s-T_{cav}$  plane  share   similarities with  the $G-T$ curves of    the AdS black holes.
	Then, we investigate  the trajectory of the light rays casted by the  black holes in a cavity.  We  reveal that  the  light trajectories   are different than   the light ray behaviors of the non rotating black holes due to the cavity effect.  Finally,  we provide evidence for the existence of  an   universal  ratio associated  with the photon sphere radius and the impact parameter. Precisely, we find  the optical  ratio   $\frac{b_{sp}}{r_{sp}} \sim \sqrt 3.$	
	\\
This paper is organized  as follows. In  section 2, we present a concise review on the    black holes surrounded
by  a cavity.  In section 3, we approach  the black hole shadows in terms of the horizon radius.  Section   4 is devoted  to   the thermal behaviors of such shadows.   In section 5, we discuss  the trajectories  of the light rays  casted by the  black holes in a cavity. Finally,  section 6   ends up with
certain  conclusions and  open questions.

\section{ Black holes in a  cavity }
In this section, we present a concise review on  the black holes in a  cavity background. In particular, we give the associated  relevant concepts.  It is recalled that in  asymptotically AdS geometries, the black holes can be  considered  thermodynamically  stable\cite{a30,a34}. In this way,  the AdS boundary serves as a reflective wall.  Alternatively,   various  works have  reported  that the black holes enclosed by a cavity can be also  thermally stable.  It has been suggested that  the   cavity   presence  could  overcome  the thermodynamics  stability problem   of the  black holes in asymptotically flat spaces. These activities push one to inspect other aspects including the optical ones.  This could unveil certain interesting behaviors by considering  static and spheric metric  structures of the black holes in a cavity physical system.  In this way, the   4-dimensional metric solution can be expressed as   follows 
\begin{equation}
\label{ }
ds^2=-f(r)dt^2+\frac{dr^2}{f(r)}+r^2(d\theta^2+\sin^2\theta d\phi^2),
\end{equation}
where $f(r)$ is the  function metric of  a black hole  in a cavity. In terms of the involved parameters \cite{a34},  this metric function   takes the following form
\begin{equation}
\label{ }
f(r)=(1-\frac{r_+}{r})(1-\frac{Q^2}{r\,r_+}).
\end{equation}
In this relation, $r_+$ indicates the horizon radius given by 
\begin{equation}
\label{ }
r_+=m+\sqrt{m^2-Q^2},
\end{equation}
 where  $m$  and $Q$  are  the charge  and the mass parameters of  the black holes, respectively.  For $Q=0$, however,  we recover the Schwarzschild solution \cite{a35}.   A close examination shows that  many models could be approached depending on the observer positions. For simplicity reasons, however, we could place the observer inside  the cavity surrounded region. This situation  can be ensured by the following  constraint
  \begin{numcases}{}
\label{con1}
   r_{cav}>r_{+} & 
   \\
   \label{con2}
   r_{cav}>r_{ob}, &
   \end{numcases}
   being  interpreted   as thermodynamic and optical   conditions,  respectively.  In Fig.(\ref{F1}), we  illustrate  the representation of   a black hole in  a  cavity  with the above  requirements. 
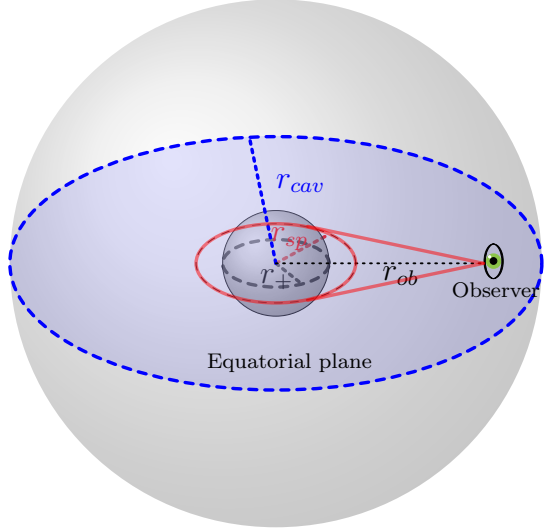
\begin{figure}[ht!]
\begin{center}
\begin{tikzpicture}[scale=0.7,line cap=round, line join=round]

\draw [rotate around={0.:(0.,0.)},dashed, line width=0.45mm] (0,0) ellipse (1cm and 0.45cm);
   \draw [rotate around={0.:(0.,0.)},dash pattern=on 3pt off 3pt, color=black] (0,0) ellipse (1.5cm and 0.75cm);
     \draw [dotted, line width=0.45mm,color=red ] (0,0)-- (1,0.55);
       \draw [dotted, line width=0.45mm ] (0,0)-- (0.5,-0.4);
  \draw(0,0) circle (1cm);
      \shade[ball color = lightgray,
    opacity = 0.5
] (0,0,0) circle (1cm);
 \draw (-0.5,0.1) node[anchor=north west] {$r_{+}$};
   \draw (-0.3,0.89) node[anchor=north west,color=red] {$r_{sp}$};
     \shade[ball color = lightgray,
    opacity = 0.3
] (0,0,0) circle (5cm);
\draw [dashed, line width=0.45mm,color=blue] (0,0) ellipse (5cm and 2.4cm);
\draw [dotted, line width=0.45mm, color=blue ] (0,0)-- (-0.5,2.4);
  \draw (-0.2,1.9) node[anchor=north west, color=blue] {$r_{cav}$};

 \draw [color=blue,fill, opacity=0.1] (0,0) ellipse (5cm and 2.4cm);
  \draw (3.7,-0.6) node[anchor=north west, rotate=90] {\frontaleye};
   \draw (-1.5,-1.5) node[anchor=north west] {{\scriptsize{Equatorial  plane}}};
   \draw (3.1,-0.15) node[anchor=north west] {\scriptsize{Observer}};
   \draw (1.8,0.15) node[anchor=north west] {$r_{ob}$};
       \draw [dotted, line width=0.25mm ] (0,0)-- (3.9,0);

 \draw [line width=0.45mm, color=red,  opacity = 0.5 ] (3.9,0)-- (0.6,0.7);
  \draw [line width=0.45mm, color=red,opacity = 0.5 ] (3.9,0)-- (0.6,-0.7);
    \draw [line width=0.45mm, color=red, opacity = 0.5] (0,0) ellipse (1.5cm and 0.75cm);

   \end{tikzpicture}
\caption{ Illustrating black hole in a cavity.}
\label{F1}
\end{center}
\end{figure} 
 
The red line represents the light transmitting to the observer placed at $r_{ob}$ being  smaller  to the  cavity radius  $r_{cav}$.   However, the blue surface  corresponds to   the equatorial plane.

\section{Shadow behaviors  of the  black holes in a  cavity system}
In this section, we investigate the optical properties of the black hole in a cavity. In particular, we study the shadow behaviors by varying the involved black hole  parameters. Precisely, we approach the shadow geometrical configurations in terms of the  one-dimensional real closed curves obtained from the equation of motion\cite{a7,a16,a19}. The  geometric behaviors of such curves are controlled by $r_+$ and $Q$ parameters. To start, we consider  the Hamilton-Jacobi equation
\begin{equation}
\frac{\partial S}{\partial \tau }=-\frac{1}{2}g^{\alpha\beta}p_{\alpha}p_{\beta},
\end{equation}%
where $S$ and $\lambda $  are  the Jacobi action and the affine parameter along the geodesics, respectively.  $p^\alpha$ represents  the conjugate momentum of the  black hole in a cavity system. In the spherically symmetric spacetime,  the Hamiltonian  which describes  the  photon motion can be  expressed as  follows
\begin{equation}
H=\frac{1}{2}g^{\alpha\beta}p_{\alpha}p_{\beta}=0.  \label{EqHamiltonian}
\end{equation}
 Considering a particular  motion of the photons  in  the equatorial plane  $\theta =\frac{\pi }{2}$,  the  Hamiltonian  equation  reduces  to the following form \begin{equation}
(rf(r)p_{r})^2-r^2p_{t}^{2}+f(r)p_{t}^{2}=0.
\label{EqNHa}
\end{equation}
Apply the Hamiltonian-Jacobi  formalism, the equations describing   the motion of  the photons  can be formulated as
\begin{eqnarray}
\frac{dt }{d\lambda}&=&\frac{E}{f(r)}  \notag \\
&&  \notag \\
\frac{dr }{d\lambda} &=&\pm\sqrt{f(r)\left(\frac{E^{2}}{f(r)}- \frac{%
L^{2}}{r^{2}}\right) }  \notag \\
&&  \notag \\
\frac{d\phi }{d\lambda} &=&-\frac{L}{r^{2}}
\label{Eqmotion}
\end{eqnarray}%
where $E=-p_t$ and $L=p_\phi$ are the conserved total energy  and the conserved angular momentum of the photon, respectively.   It turns out that the geometric shape of a black hole   can be  completely described  by the limit of their  shadows being the visible shape of the unstable  closed geodesic curves  of the photons.  To approach  such  behaviors, one  can exploit the radial equation of motion given by 
\begin{equation}
\label{22}
\Big(\frac{dr}{d\tau}\Big)^2+V_{eff}(r)=0,
\end{equation}
where $V_{eff}(r)$ indicates the effective potential for a  radial particle motion  in the space-time. In particular, it reads as 
\begin{equation}
\label{23}
V_{eff}=f(r)\left( \frac{L^{2}}{r^{2}}-\frac{E^{2}}{f(r)}\right).
\end{equation}
An examination reveals that the maximal  value of the effective potential  corresponds   to  the radius  of the  circular orbits $r_{sp}$. To get the radius of  photon unstable  circles, one  should   consider  the following  constraint
\begin{equation}
\label{24}
V_{eff}=\frac{d V_{eff}}{d r}\Big|_{r=r_{sp}}=0.
\end{equation}
Using  Eq.\eqref{23}  and  Eq.\eqref{24}, one can obtain 
\begin{equation}
\label{25}
V_{eff}|_{r=r_{sp}}=\frac{d V_{eff}}{d r}\Big|_{r=r_{sp}}= \left\{
    \begin{array}{ll}
        &f(r_{sp})\left( \frac{L^{2}}{r_{sp}^{2}}-\frac{E^{2}}{f(r_{sp})}\right)=0, \\
        \\
        &L^2\left(\frac{r_{sp}f'(r_{sp})-2f(r_{sp})}{r_{sp}^3}\right)=0,
    \end{array}
\right.
\end{equation}
where the  notation $f'(r)=\frac{\partial f(r)}{\partial r}$ has been considered. By the help of the Eq.(\ref{25}), we can obtain  the equation  constraint  concerning  the   radius  of the unstable photons sphere  via the relation \begin{equation}
\label{26}
r_{sp}f'(r_{sp})-2f(r_{sp})=0.
\end{equation}
The  real and  the positive solution of this equation is given by 
\begin{equation}
\label{Ca2}
r_{sp}=\frac{3 (Q^2+ r_+^2)+\sqrt{9 Q^4-14 Q^2 r_+^2+9 r_+^4}}{4  r_+}.
\end{equation}
It has been remarked that the  photon sphere radius $r_{sp}$ which  depends on  the charge $Q$ and the horizon radius $r_+$ can recover certain previous results.  Taking  $Q=0$ and $r_+=2M$, we obtain the radius of the unstable photon sphere of  the  Schwarzschild black hole solution\cite{a7,a9,a16}.  By the help of  the  radial and the  angular geodesic equations, the orbit equation for the photon reads as
\begin{equation}
\frac{dr}{d\phi}=\pm \frac{r^{2}}{L}\sqrt{f(r)\left(\frac{E^{2}}{f(r)}- \frac{%
L^{2}}{r^{2}}\right) }.
\label{Eqorbit}
\end{equation}
The photon orbit is constrained  by
\begin{equation}
\label{Ca1}
 \frac{dr}{d\phi}\Big\vert_{r=r_{sp}}=0.
 \end{equation}
Using the Eq.(\ref{Ca1}),  the previous equation take the following form 
\begin{equation}
\frac{dr}{d\phi}=\pm r\sqrt{f(r)\left[\frac{r^{2}f(r_{sp})}{r_{sp}^{2}f(r)} -1\right] }%
.  \label{EqTp}
\end{equation}
Considering the  light ray sent  from a static observer situated at $r_{ob}$ inside the  cavity  and transmitted  into the past with an angle $\alpha_{ob}$,   we have 
\begin{equation}
\cot\alpha_{ob}=\frac{\sqrt{g_{rr}}}{\sqrt{g_{\phi\phi}}}\frac{dr}{d\phi}{\Big{|}}_{r=r_{ob}}=\frac{1}{r\sqrt{f(r)}}\frac{dr}{d\phi}{\Big{|}}_{r=r_{ob}}.
\label{ff1}
\end{equation}
Exploiting Eq.\eqref{ff1},   one obtain  the angle of the observer as a function of the various parameters 
\begin{equation}
 \sin^{2}\alpha_{ob}=\frac{f(r_{ob})r_{sp}^{2}}{r_{ob}^{2}f(r_{sp})}.
\label{effph}
\end{equation}
In this context,  one can   get   the  angular radius of the black hole shadows as a function of    the circular orbit radius of the photon appearing in   Eq.\eqref{Ca2}. Precisely,  the  shadow radius of the black hole observed by a static observer placed  at $r_{ob}$ is given by 
\begin{equation}\label{shara}
r_{s}=r_{ob}\sin\alpha_{ob}=\left. R\sqrt{\frac{f(r_{ob})}{f(R)}}\right|_{R=r_{sp}}.
\end{equation}
According to  \cite{a36},  the   apparent shape of  the black hole   shadow in a cavity system   can be  obtained by using the celestial
coordinates $x$ and $y$  which can be expressed as 
 \begin{eqnarray}
x &=&\lim_{r_{0}\longrightarrow \infty }\left( -r_{0}^{2}\sin \theta _{0}%
\frac{d\phi }{dr}\Big\vert_{(r_{0},\theta _{0})}\right) ,  \notag \\
&&  \notag \\
y &=&\lim_{r_{0}\longrightarrow \infty }\left( r_{0}^{2}\frac{d\theta }{dr}%
\Big\vert_{(r_{0},\theta _{0})}\right).
\end{eqnarray}
In Fig(\ref{f220}), we plot the shadow geometry  for different values of the charge $Q$.
\begin{figure}[!ht]
		\begin{center}
		\centering
			\begin{tabbing}
			\centering
			\hspace{8.cm}\=\kill
			\includegraphics[scale=0.8]{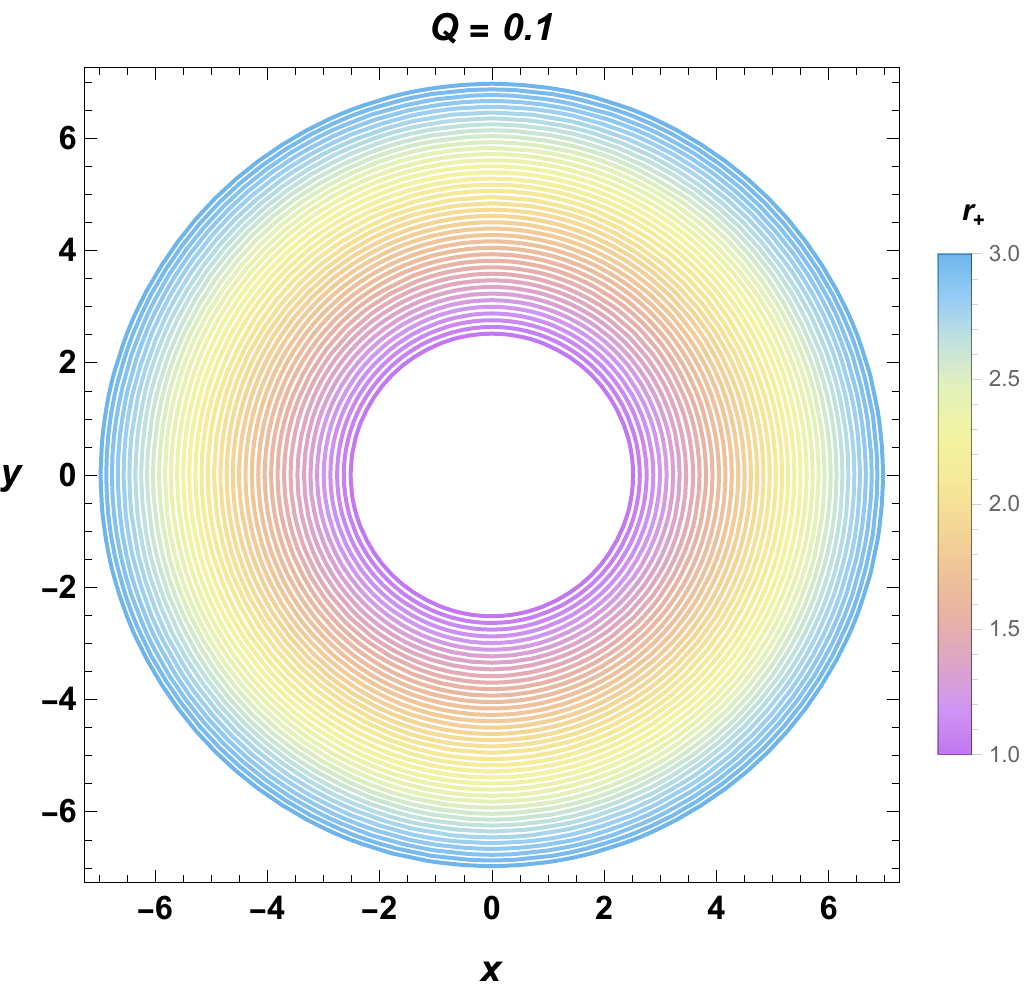} 
	\hspace{0.1cm}		\includegraphics[scale=0.8]{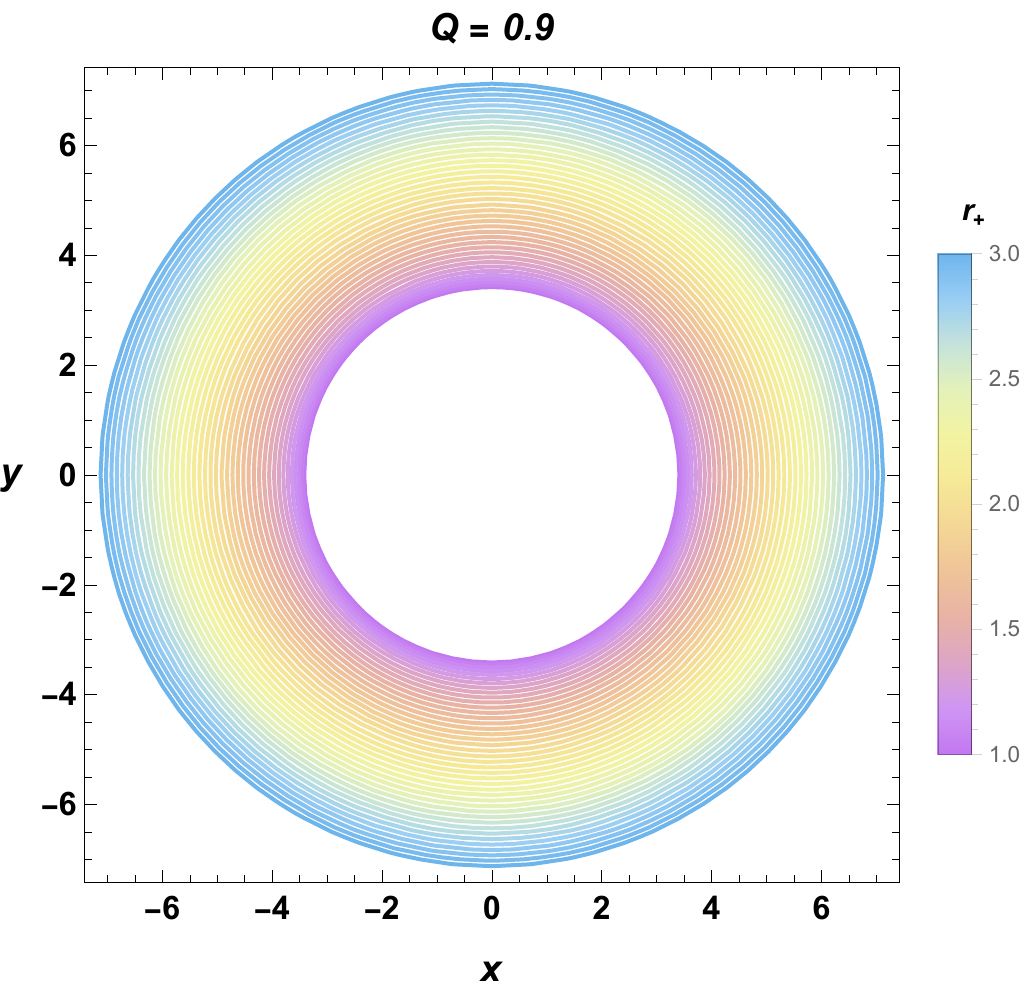}\\ 
	
		   \end{tabbing}
\caption{ \it \footnotesize  Shadow variations of  a black hole in a cavity   system  as a function of $r_+$  for fixed values of $Q$ by placing the  observer in the equatorial plane and taking    $r_{ob}=15$ and   $r_{cav}=20$.} 
\label{f220}
\end{center}
\end{figure}
As expected,  the geometry of the shadows is a perfect  circular due to the absence of the rotating parameter.   It has been remarked that the charge of the black hole in a cavity is a relevant parameter controlling the optical  behaviors including the shadow size. Indeed, the   latter  increases with  the horizon radius.  Taking a  fixed   value of $r_+$,  the shadow size increases by increasing the charge $Q$.  It is denoted that for $Q=0$,  we recover the shadow of  the Schwarzschild black hole\cite{a7,a9,a16}. To inspect the shadow radius  variation, we  consider its variation in terms of   $r_+$   for different 
 values of the charge. This is illustrated in   Fig(\ref{f222}).
  \begin{figure}[ht!]
\begin{center}
\includegraphics[width=10cm, height=5.5cm]{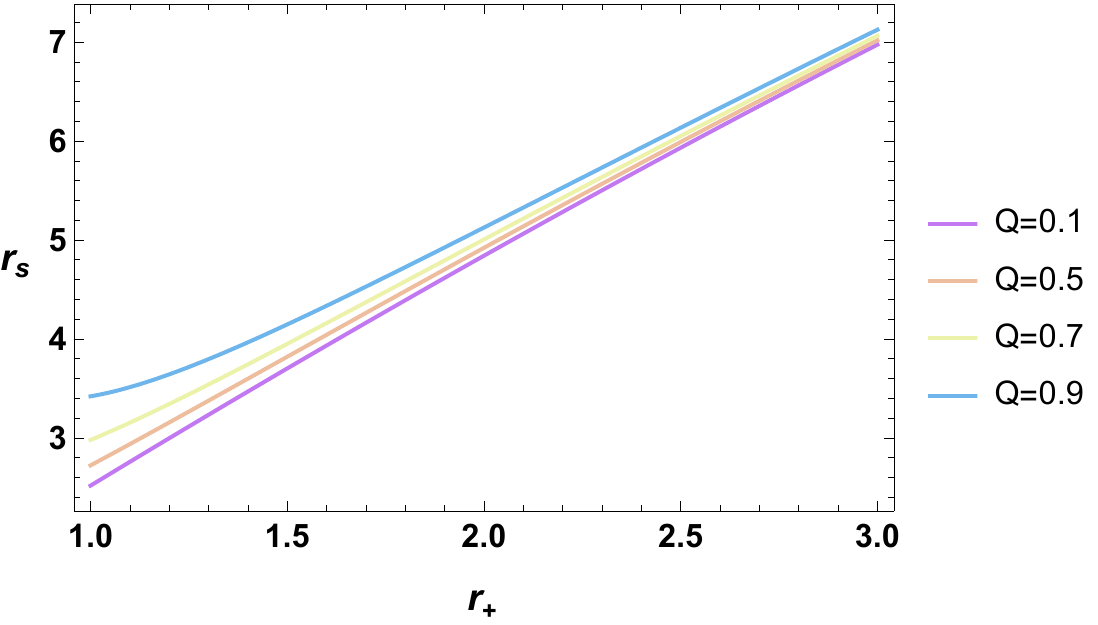}
\caption{ \it \footnotesize  Shadow radius as  a function of  $r_+$ for fixed values of the charge. The observer in the equatorial plane and  positioned in   $r_{ob}=15$, we take the  cavity radius  $r_{cav}=20$.}
\label{f222}
\end{center}
\end{figure}
This figure confirms the previous findings  where the radius augments with the charge. This   quantity has been interpreted as a geometric parameter controlling  the shadow size. 
 
\section{Shadow thermal behaviors of  black holes }
 Motivated by certain  results associated with the link  between the  black holes in  a cavity and the thermodynamics of  the AdS black holes, we study shadow behaviors by varying  the cavity  temperature. It is recalled that the temperature is a crucial  needed quantity  to approach   the stability behaviors.  Indeed,  the  Hawking temperature  is given by 
\begin{equation}
\label{Ca3}
T_H=\left.\frac{1}{4\pi}\frac{df(r)}{dr}\right|_{r=r_{+}}.
\end{equation}
In the cavity system, however,   the temperature  related  to the  cavity radius $r_{cav}$.  Concretely, it   is expressed as  follows
\begin{equation}
\label{shara}
T_{cav}=\frac{T_H}{\sqrt{f(r_{cav})}}
\end{equation}
which  can be given   in terms of the involved physical   parameters \cite{a30,A30,a34}. According to such works,   the  calculations provide 
\begin{equation}\label{shara}
T_{cav}=\frac{r_+^2-Q^2}{4\pi r_+\sqrt{f(r_{cav})}}.
 \end{equation}
It has been  remarked  that the cavity    temperature  $T_{cav}$ and the shadow geometry in a cavity are controlled by the radius horizon. Inspired by such a  link, we investigate   the shadow behaviors by varying the temperature. It has been observed that, for $r_{cav}>r_{+}$, the temperature of the cavity system is similar to the one  of the  charged AdS black hole.  The only  difference is for $r_{cav}=r_{+}$  where  the  cavity temperature   $T_{cav}$ diverges\cite{a34}.  As proposed in  Fig.(\ref{F1}),  the observer is placed inside the cavity to visualize the transmitting lights in terms of  the shadow real closed  curves. By the help of   the method reported in \cite{a36},  we study  the thermal  shadow behaviors.   In Fig.(\ref{F2}), we illustrate the shadows  as   functions of the  cavity temperature $T_{cav}$  for fixed values of  the charge $Q$. To get a concrete model, we consider a situation where one has $r_{cav}=20$ and  $r_{ob}=15$. 
\begin{figure}[!ht]
		\begin{center}
		\centering
			\begin{tabbing}
			\centering
			\hspace{8.cm}\=\kill
			\includegraphics[scale=0.8]{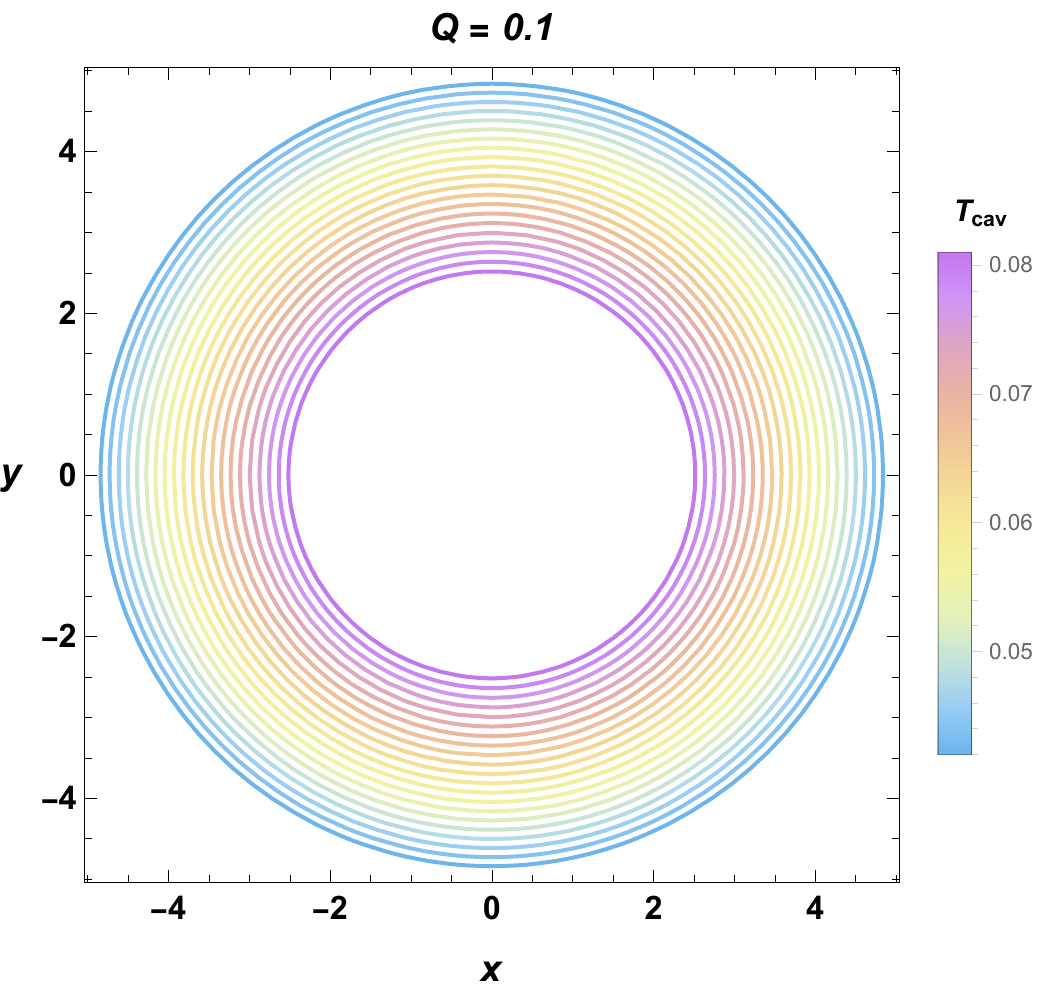} 
	\hspace{0.1cm}		\includegraphics[scale=0.8]{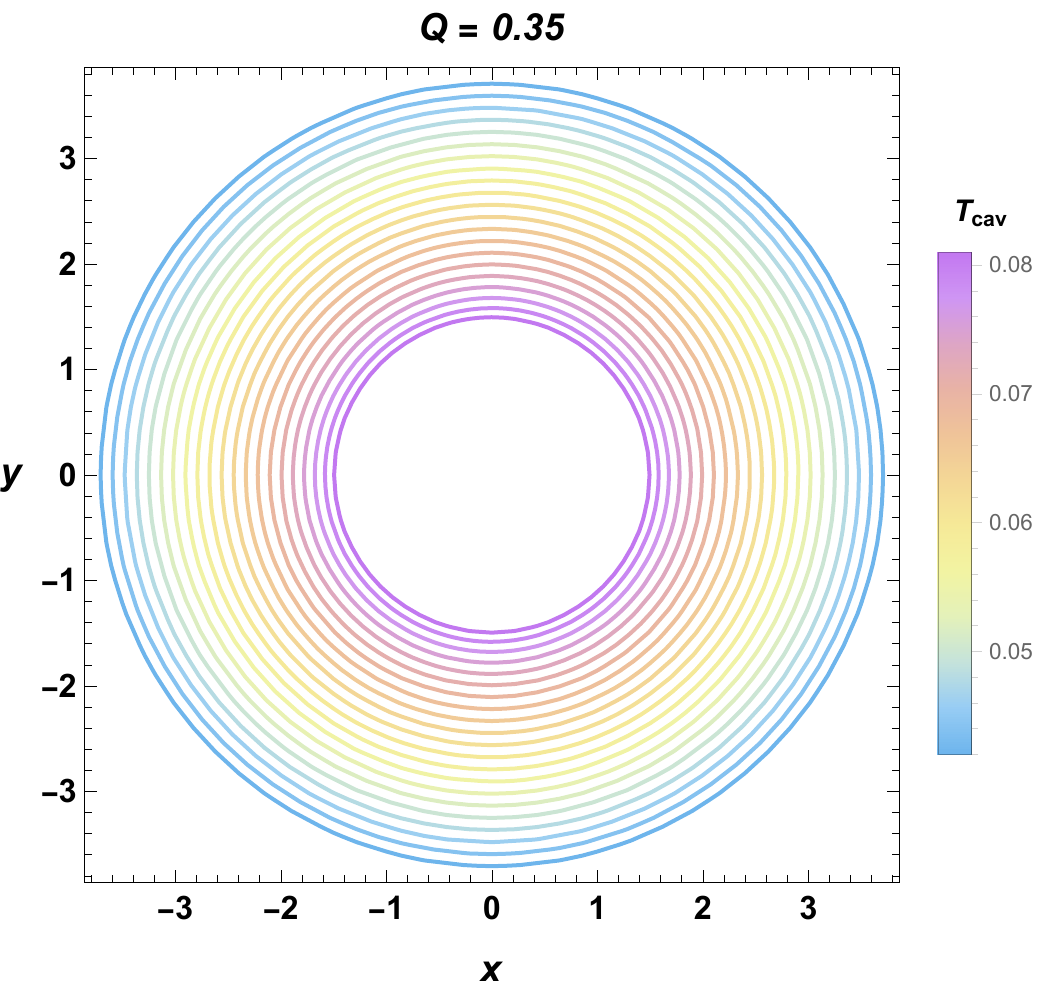}\\ 
	
		   \end{tabbing}
\caption{ \it \footnotesize  Shadow behaviors of  a black hole in a cavity   system  as a function of the cavity temperature   for fixed values of $Q$ by placing the  observer in the equatorial plane and taking    $r_{ob}=15$ and   $r_{cav}=20$.} 
\label{F2}
\end{center}
\end{figure}
In   this way, the shadow size  increases  by  decreasing the temperature of the  cavity system. It follows from this figure   that the charge $Q$ decreases the shadow radius of the  black hole in a cavity. Augmenting the charge $Q$,   we observe  that the temperature variation  is contrary to the horizon radius one  as illustrated  in Fig.(\ref{f222}).  These  inverse behaviors come from the relationship  between the cavity  temperature  and the horizon radius $r_+$.   Indeed, when  $T_{cav}$  decreases (increases),     $r_+$  increases (decreases).   Moreover,  certain properties  of the  AdS black hole solutions  associated with the  shadow radius aspect have been conserved  in the black hole inside  the cavity.   In fact, it  is an    increasing (decreasing) function of the mass (temperature) \cite{a9,a16}.   Such thermal behaviors push one to inspect other   black  hole properties needed to support the present findings.  The corresponding  energy emission rate could be investigated.  It is worth noting  that near  the  horizon  of    the black hole,  the   quantum fluctuations can  create and annihilate  pairs of  particles. In this way,   the  particles  with  positive energies  can escape through tunneling from the black hole associated with  the Hawking radiation.   In what follows, we discuss  the involved  energy emission rate in a cavity background.   For a distant observer with the above  optical and the  thermodynamical conditions,  the high energy absorption cross section  could provide data on  the  black hole shadows.   In this regard,    it has been  remarked  that  the  absorption cross section of the black hole can  oscillate
  to an  approximated  constant value $\sigma_{lim}=\pi r_s^2$.   The  energy emission rate can be written as 
\begin{equation}
\label{72}
\frac{d^2 E(\omega)}{d\omega dt}=\frac{2\pi^{3} r_s^{2}\omega^3}{e^{\frac{\omega}{T_{{H}}}}-1},
\end{equation}
where $\omega$    indicates   the emission frequency \cite{Wei:2013kza,a16,a18,a19}, and where   $T_{H}$ is the  associated  Hawking temperature.  Using the link with  the cavity temperature,  this expression takes  the following form
\begin{equation}
\label{73}
\frac{d^2 E(\omega)}{d\omega dt}=\frac{2\pi^{3} r_s^{2}\omega^3}{e^{\frac{\omega}{ {\sqrt{f(r_{cav})}}.
 T_{{cav}}}}-1}.
\end{equation}
The  energy emission rate is   illustrated   in Fig.(\ref{FA}) as a function of $\omega$ by varying $T_{cav}$ and  $Q$. 
\begin{figure}[!ht]
		\centering
			\begin{tabbing}
			\hspace{6cm}\= \hspace{4cm}\=\kill
			\includegraphics[scale=.7]{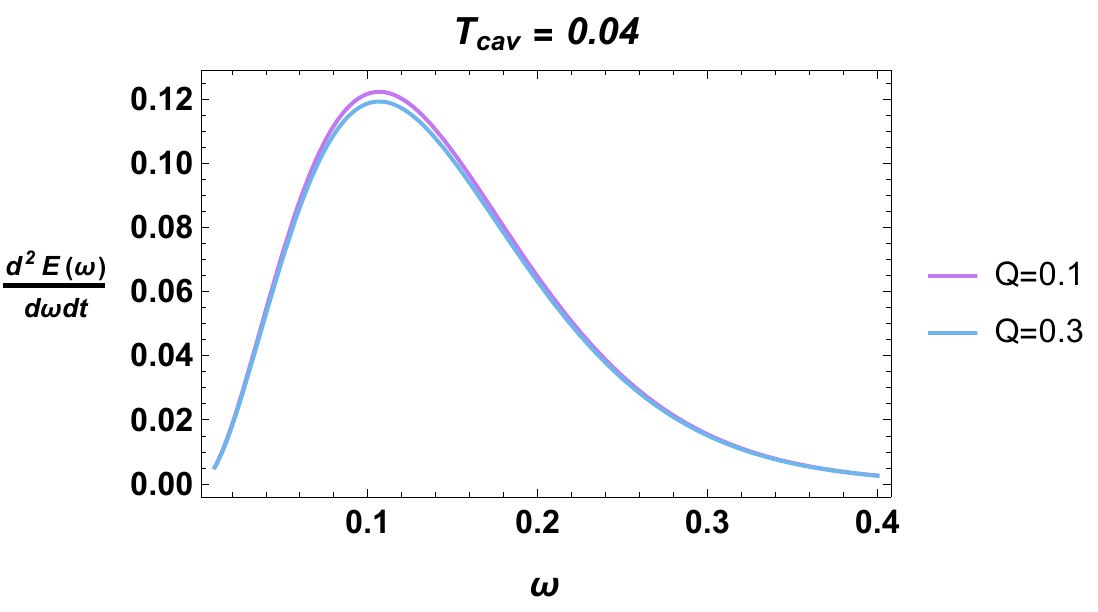} \> 
			 \hspace{2cm} \includegraphics[scale=.7]{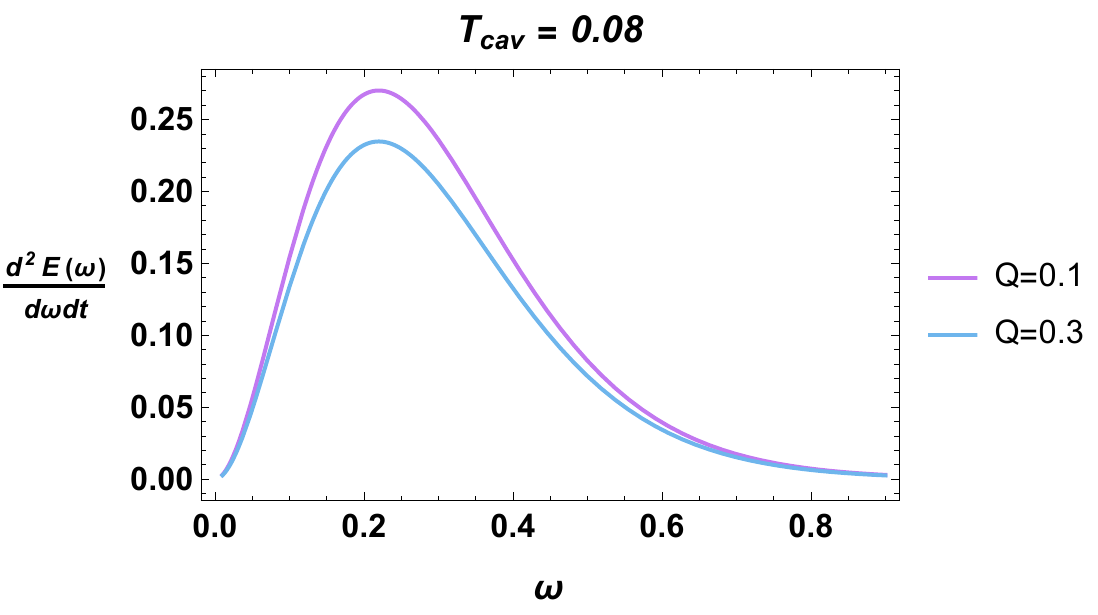} \\
		          \end{tabbing}
		          \vspace{-1cm}
\caption{{\it \footnotesize Energy emission rate for different  values of $T_{cav}$ and  $Q$,  by placing the  observer in the equatorial plane and taking    $r_{ob}=15$ and   $r_{cav}=20$.}}
\label{FA}
\end{figure}
For small values of the   cavity temperature,  it  has been observed that  the charge does not bring any relevant effect.  Augmenting  such a  temperature,   the   energy emission rate
maximum increases by decreasing the charge   $Q$.  Fixing the charge values,  this maximum increases with  $T_{cav}$.  This shows  that   $T_{cav}$ and  $Q$ involve opposite effects on the involved optical behaviors to ensure the stability. This finding confirms   the  previous  obtained results dealing   with  black holes  in cavities.\

Inspired by the  similarities between  the black hole  thermodynamics in AdS geometries and in isothermal  backrounds, we  approach the   cavity  temperature   from an optical point of view.
Indeed,  we  plot in  Fig.(\ref{F4})  the cavity  temperature  as a function of the shadow radius by  varying the charge $Q$. 
 \begin{figure*}[!ht]
		\begin{center}
		\begin{tikzpicture}[scale=0.2,text centered]
		\node[] at (-40,1){\small  \includegraphics[width=10cm, height=6.5cm]{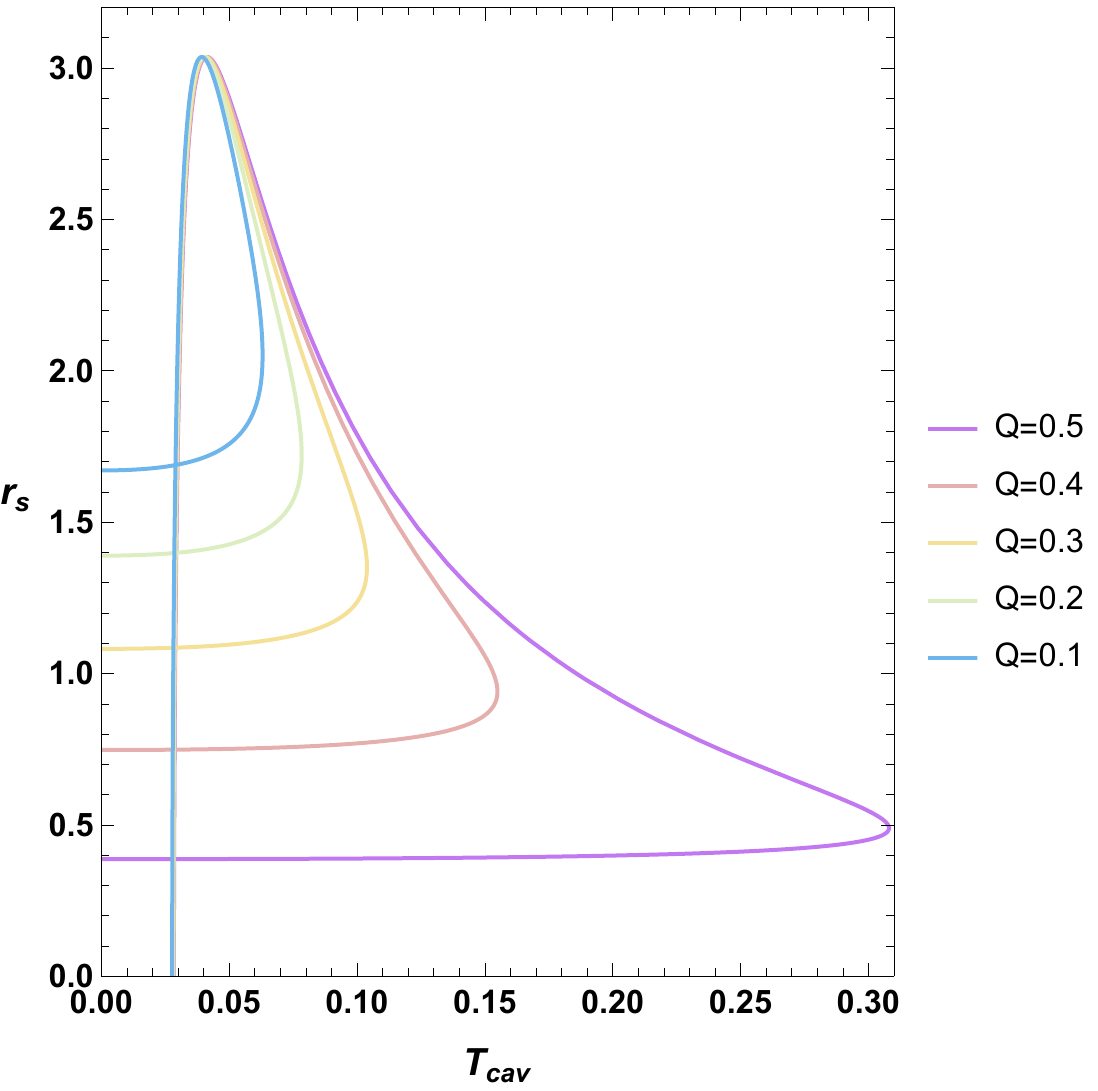}};
\fill[red] (-57.1,-11.9) circle (0.3);
\node[color=red] at (-57.1,-15){\scriptsize {$T_{HP}=0.0284$}};

\end{tikzpicture}	
\caption{{\it \footnotesize  The behaviors of the shadow radius in terms of the cavity temperature. In particular, we take  the observer in the equatorial plane and positioned in $r_{ob} = 3$ and the cavity radius $r_{cav}=20$.}}
\label{F4}
\end{center}
\end{figure*}
The shadow radius increases  by decreasing $T_{cav}$ and  decreases   with  the charge $Q$.   This behavior  is  confirmed  in Fig.({\ref{F4}}),  by using the  optical and the thermodynamical conditions given by the equations Eq.(\ref{con1}) and  Eq.(\ref{con2}) for   $r_{ob}=3$ associated with  an  observer placed   closed  to the sphere photon radius of the black hole in a  cavity.  An examination shows that the  curves   in  the $r_s-T_{cav}$ plane   share   similarities with  the $G-T$ curves of    the charged and the  rotating AdS black holes, where $G$ and $T$  are the Gibbs free energy and the temperature, respectively.  To unveil such similar behaviors, we should exploit  such thermodynamical quantities.   According to \cite{a34}, the  Gibbs free energy expression  of a black hole in a cavity  takes the following  form 
\begin{equation}
G=r_{cav}\left[\f{7\f{r_+}{r_{cav}}+\f{Q^2}{r_+r_{cav}}-8}{12\sqrt{f(r_{cav})}}+\f23\rt]. \label{G}
\end{equation}
  Solving the equation $G=0$, we can get the value of the  phase transition temperature $T_{HP}$. Varying the charge $Q$ parameter, this temperature is found to be constant
  \begin{equation}
\label{ }
T_{HP}= 0.0284.
\end{equation}
 Moreover,  it   has been observed that the $r_s-T_{cav}$ curves involve   the  swallowtail behavior  observed in the $G-T$ plane. Indeed, the   $T_{HP}$ temperature in the   $G-T$ plane      coincides with the    $T_{HP}$ temperature in  the $r_s-T_{cav}$ plane. This result could be used to  confirm the relation between the   thermodynamics    and the optical aspects of the black holes in a cavity. As expected, certain thermal  quantities  of  the black holes including the phase transition temperature $T_{HP}$ can be approached using  the shadow findings.
 %It is interesting to examine such a similarity.  In such AdS black holes, these geometries configurations are with phase transitions. 

%{ \bf To be completed in good ways.}
 \section{Trajectories  of the light rays by black holes in a cavity }
%Properly, wen the black hole is illuminated by a source, therefore is capable to send an image due to the particular trajectories followed by the light emitted in its vicinity.
 In this section, we study the  trajectory of the light rays casted   by  the black holes  in a cavity.  In particular, we determine   the  light trajectory around  the  black hole in a cavity by varying the cavity temperature $T_{cav}$. The light  trajectory  casted by  the black holes  in cavities can be established by using  numerical computations  associated with  the following  orbit equation 
\begin{equation}
\frac{du}{d\phi}= \sqrt{\frac{1}{b^2}-u^2(1-ur_+)(1-\frac{Q^2u}{r_+}) },
\label{Eqorbit1}
\end{equation}
where  one has used   the change variable $u=\frac{1}{r}$ and where $b$ is the so-called the impact parameter given by $b=\frac{\lvert L \rvert}{E}$\cite{a37,a38}. Using Eq(\ref{Eqorbit1}),  we can solve $\phi$ with respect to $u$  in order to depict  the trajectory of the light ray in a cavity   background. To approach the  trajectory of the light rays casted  by the  black holes  in a such system, we need to find   the regions  corresponding to  the light ray  trajectory possibilities. Indeed, they can be determined   by the help of the effective potential given by  Eq.(\ref{23}). In  Fig.(\ref{aa1b}), we plot such an effective potential as  a function of the radial coordinate $r$ for different values of  the cavity temperature $T_{cav}$ and the charge $Q$.  
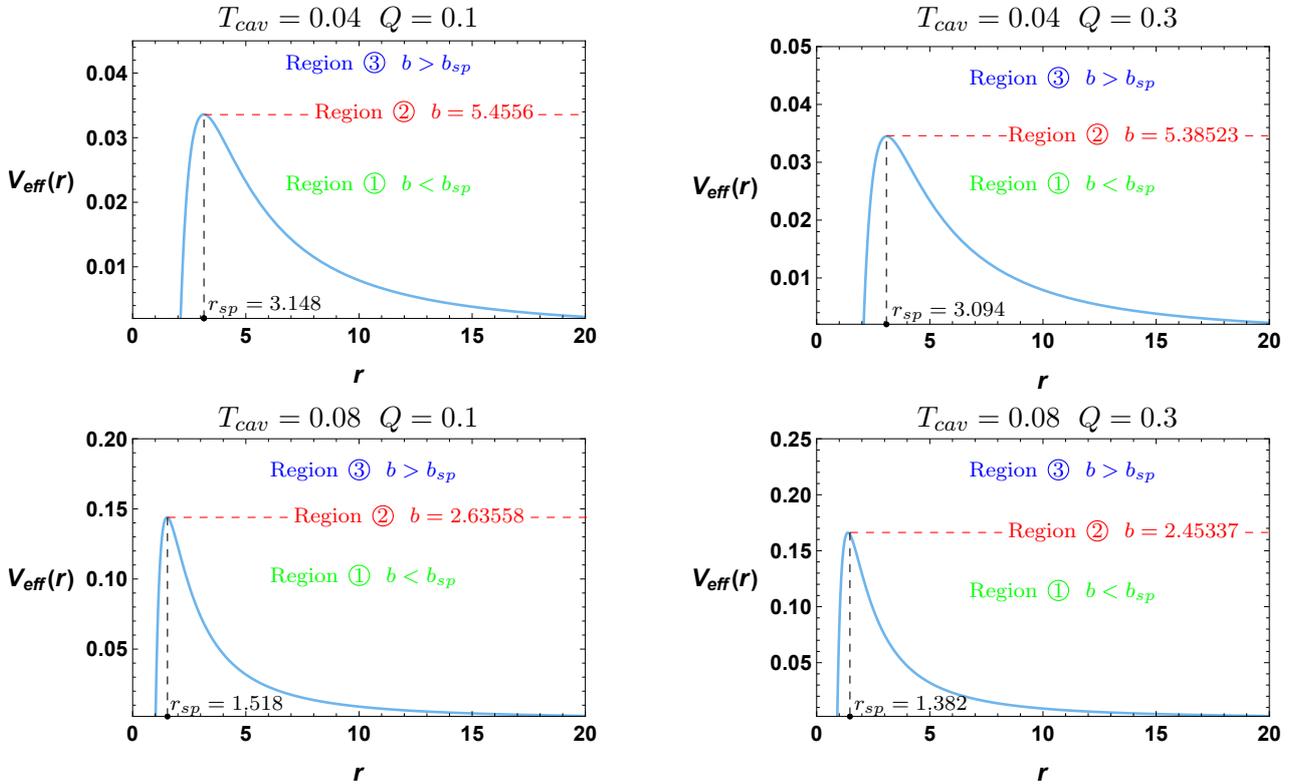
\begin{figure*}[!ht]
		\begin{center}
		\begin{tikzpicture}[scale=0.2,text centered]
		\hspace{-1.2 cm}
\hspace{0.7 cm}\node[] at (-40,1){\small  \includegraphics[scale=0.85]{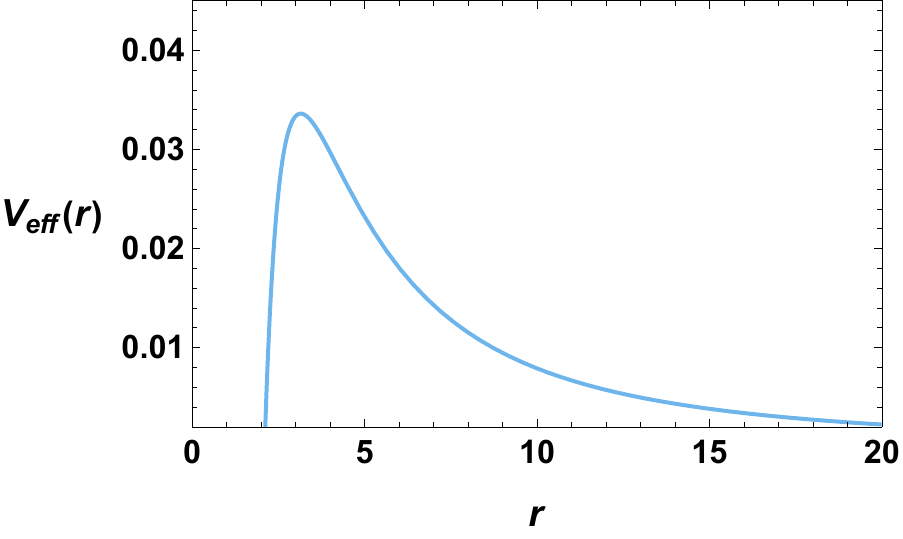}};
\node[] at (5,1){\small  \includegraphics[scale=0.85]{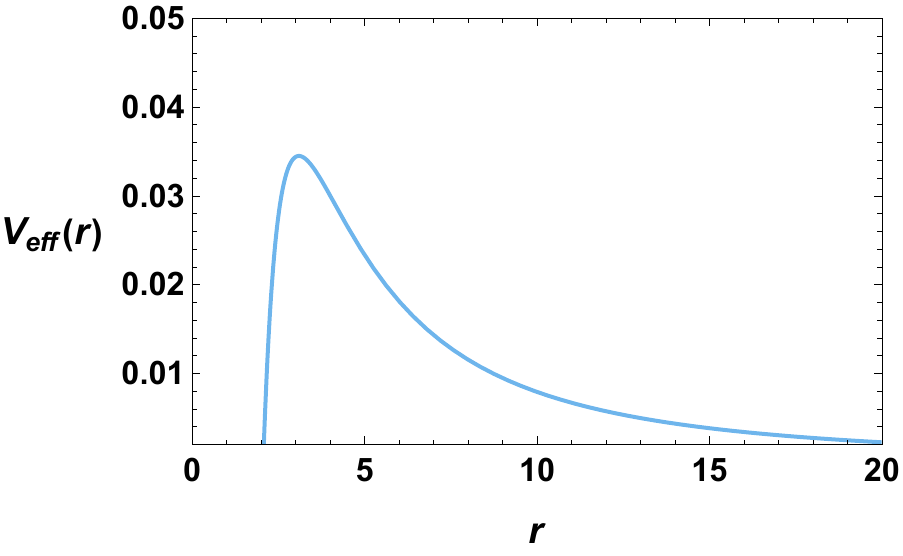}};
\hspace{0.0 cm}\node[] at (-40,-25){\small  \includegraphics[scale=0.85]{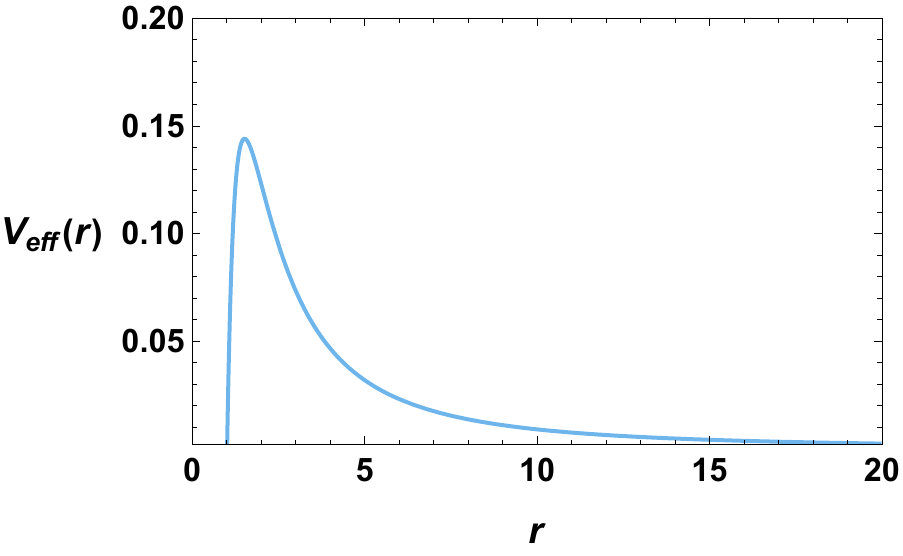}};
\node[] at (5,-25){\small  \includegraphics[scale=0.85]{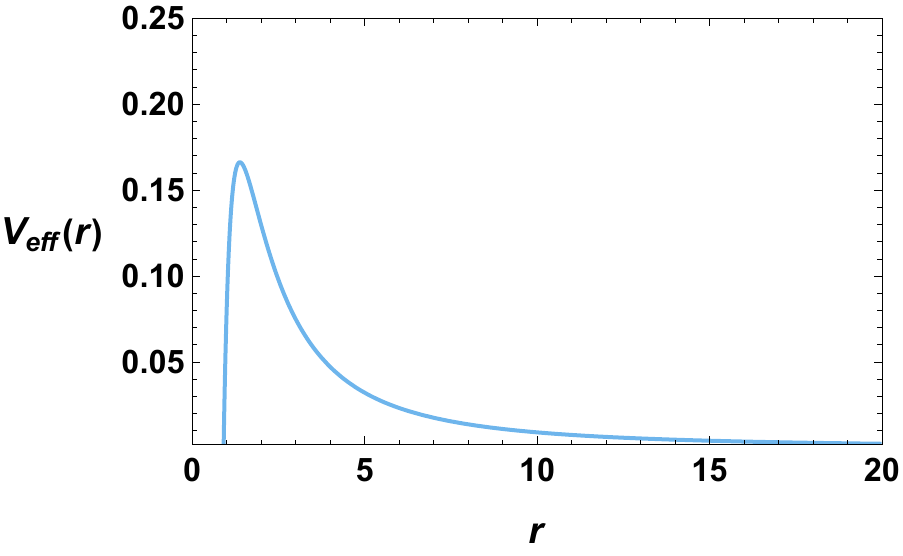}};
\draw[dashed] (-1.6,-6.2) -- (-1.6,6.4);
\draw[dashed,color=red] (-1.6,6.3) -- (6.5,6.3);
\draw[dashed,color=red] (22,6.3) -- (23.5,6.3);
\draw[dashed] (-46.5,-5.8) -- (-46.5,7.7);
\draw[dashed,color=red] (-46.5,7.7) -- (-39.5,7.7);
\draw[dashed,color=red] (-24.5,7.7) -- (-21.8,7.7);

\draw[dashed,color=red] (-4,-20) -- (6.5,-20);
\draw[dashed,color=red] (22,-20) -- (23.5,-20);
\draw[dashed] (-4,-20) -- (-4,-32);

\draw[dashed] (-48.9,-19) -- (-48.9,-32);
\draw[dashed,color=red] (-48.9,-19) -- (-40.9,-19);
\draw[dashed,color=red] (-25,-19) -- (-20.8,-19);
\node[color=black] at (-36,-16){\scriptsize {\color{blue}{Region \let\textcircled=\pgftextcircled\textcircled{3}}\; $b>b_{sp}$}};
\node[color=black] at (-33,-19){\scriptsize {\color{red}{Region \let\textcircled=\pgftextcircled\textcircled{2}}\; $b=2.63558$}};
\node[color=black] at (-36,-23){\scriptsize {\color{green}{Region \let\textcircled=\pgftextcircled\textcircled{1}}\; $b<b_{sp}$}};
\fill (-48.9,-32.2) circle (0.2);
\node[color=black] at (-45,-31.5){\scriptsize {$r_{sp}=1.518$}};

\node[color=black] at (-37,14){\small {$T_{cav} = 0.04 $}\; {$Q = 0.1 $}};
\node[color=black] at (-37,-12.5){\small {$T_{cav} = 0.08 $}\; {$Q = 0.1 $}};
\node[color=black] at (9,14){\small {$T_{cav} = 0.04 $}\; {$Q = 0.3 $}};
\node[color=black] at (9,-12.5){\small {$T_{cav} = 0.08 $}\; {$Q = 0.3 $}};
\node[color=black] at (-35,11){\scriptsize {\color{blue}{Region \let\textcircled=\pgftextcircled\textcircled{3}}\; $b>b_{sp}$}};
\node[color=black] at (-32,7.8){\scriptsize {\color{red}{Region \let\textcircled=\pgftextcircled\textcircled{2}}\; $b=5.4556$}};
\node[color=black] at (-35,3){\scriptsize {\color{green}{Region \let\textcircled=\pgftextcircled\textcircled{1}}\; $b<b_{sp}$}};

\node[color=black] at (10,10){\scriptsize {\color{blue}{Region \let\textcircled=\pgftextcircled\textcircled{3}}\; $b>b_{sp}$}};
\node[color=black] at (14,6.3){\scriptsize {\color{red}{Region \let\textcircled=\pgftextcircled\textcircled{2}}\; $b=5.38523$}};
\node[color=black] at (10,3){\scriptsize {\color{green}{Region \let\textcircled=\pgftextcircled\textcircled{1}}\; $b<b_{sp}$}};
\fill (-1.6,-6.2) circle (0.2);
\node[color=black] at (2.5,-5.4){\scriptsize {$r_{sp}=3.094$}};
\fill (-46.5,-5.8) circle (0.2);
\node[color=black] at (-42.5,-5){\scriptsize {$r_{sp}=3.148$}};
\fill (-46.5,-5.8) circle (0.2);
\node[color=black] at (10,-16){\scriptsize {\color{blue}{Region \let\textcircled=\pgftextcircled\textcircled{3}}\; $b>b_{sp}$}};
\node[color=black] at (14,-20){\scriptsize {\color{red}{Region \let\textcircled=\pgftextcircled\textcircled{2}}\; $b=2.45337$}};
\node[color=black] at (10,-24){\scriptsize {\color{green}{Region \let\textcircled=\pgftextcircled\textcircled{1}}\; $b<b_{sp}$}};
\fill (-4,-32.2) circle (0.2);
\node[color=black] at (0.,-31.5){\scriptsize {$r_{sp}=1.382$}};
\end{tikzpicture}	
\caption{{\it \footnotesize  The effective potential behaviors for different values of the charges and the cavity temperature,  by  taking the  cavity radius  $r_{cav}=20$.}}
\label{aa1b}
\end{center}
\end{figure*}

 This potential increases and reaches a maximum at the photon sphere associated with    $b_{sp}$  representing  the impact parameter of   the spinning  light rays.  This quantity verifies the following constraint 
\begin{equation}
\label{ }
V_{eff}(r_{sp})=\frac{1}{b^2_{sp}}.
\end{equation} 
Two values of  $T_{cav}$ will  been dealt with, being  $T_{cav}=0.04, 0.08$.   For $Q=0.3$, they provide two   impact parameter values $b_{sp}=5.39$ and  $b_{sp}=2.46$,  as   shown   in  Fig.(\ref{aa1b}), corresponding  to the  photon sphere radius  $r_{sp}=3.09$ and  $r_{sp}=1.38$, respectively.    For $Q=0.1$,  however, 
 the corresponding  impact parameter and the photon sphere radius  increase.  Fixing the charge values,  the  effective potential increases  by decreasing   the cavity temperature. We observe that the photon sphere radius $r_{sp}$  also  decreases  by increasing the cavity temperature.  It has been remarked  that the impact parameter value  $b_{sp}$ provides    the trajectories  of the light rays  in three  different regions. These  regions are denoted by region \let\textcircled=\pgftextcircled\textcircled{1}, region \let\textcircled=\pgftextcircled\textcircled{2} and region \let\textcircled=\pgftextcircled\textcircled{3} corresponding to $b<b_{sp}$, $b=b_{sp}$ and $b>b_{sp}$, respectively.
 
In the region  \let\textcircled=\pgftextcircled\textcircled{1},  the light ray falls  into the black hole in a cavity due to the values of  the impact parameter lower to $b_{sp}$.  In region \let\textcircled=\pgftextcircled\textcircled{3},  however, the light rays near the black hole in a cavity system  are reflected back.   In the  region   \let\textcircled=\pgftextcircled\textcircled{2},   the light ray comes into the photon sphere making  an infinite  turn number  around the black  hole  due to the non vanishing value of  the angular velocity. The associated  orbit is circular and unstable. To illustrate these regions,  we plot in Fig.(\ref{aa}) the trajectories of the light rays in the polar coordinates $(r,\phi)$ for different values of  the cavity temperature $T_{cav}$ and the charge $Q$. To analyse the effect of the cavity temperature in the  light ray trajectories,  we vary the impact parameter $b$ in the range $]0,7]$. The step between  two values of   the impact parameter is 1/10 for all light rays. 
\begin{figure*}[!ht]
		\begin{center}
		\begin{tikzpicture}[scale=0.2,text centered]
		\hspace{-1.2 cm}
\node[] at (-40,1){\small  \includegraphics[scale=0.40]{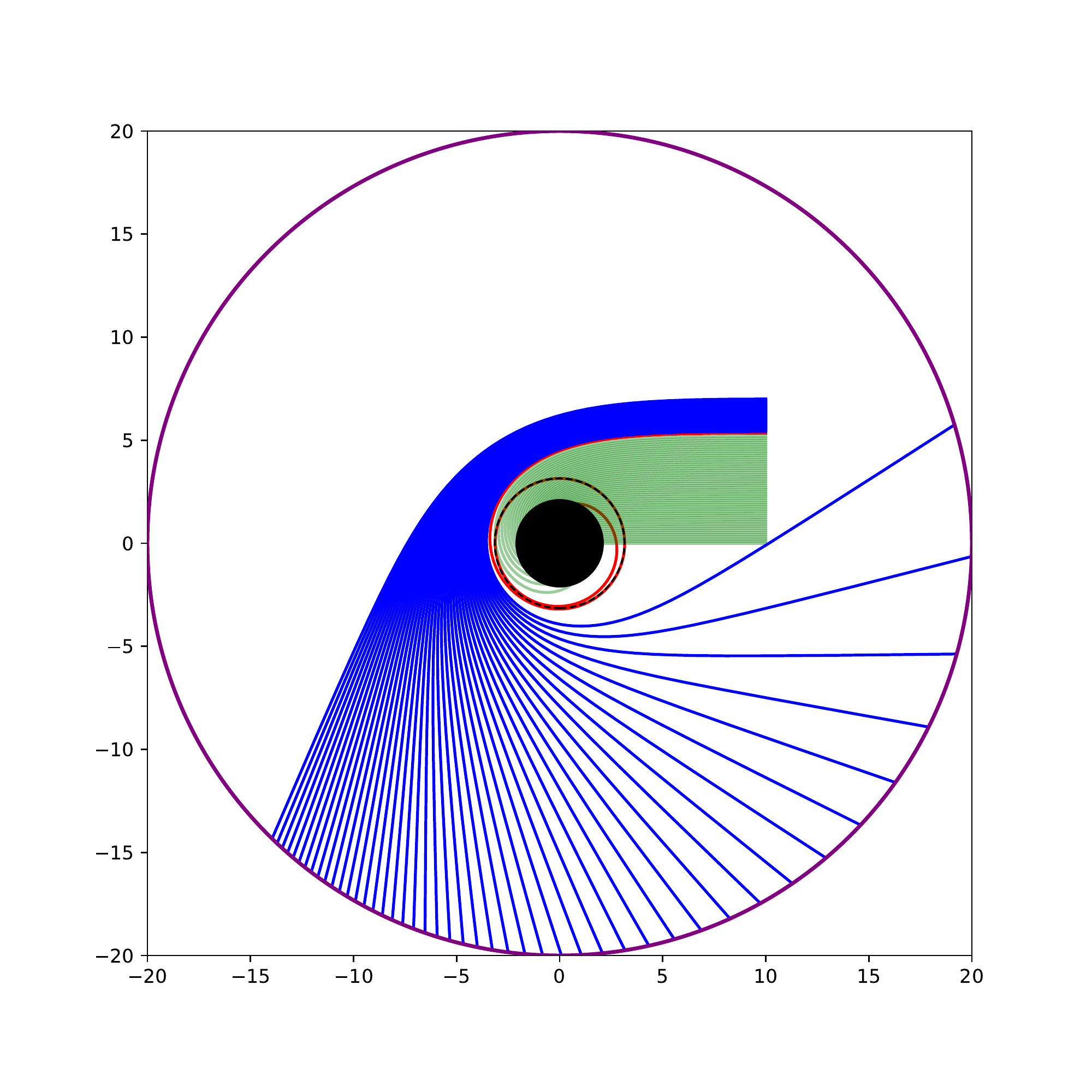}};
\node[] at (10,1){\small  \includegraphics[scale=0.40]{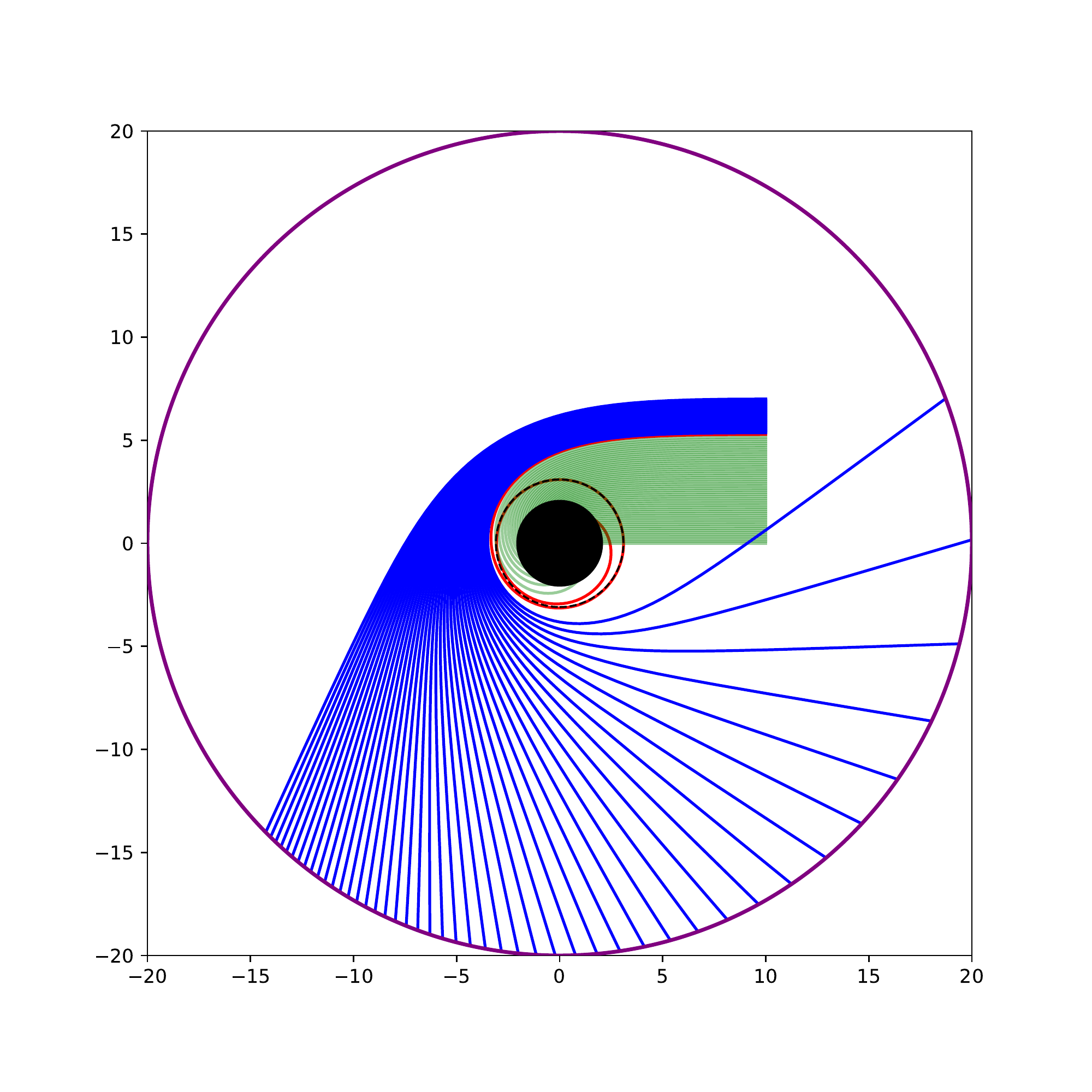}};
\node[] at (-40,-40){\small  \includegraphics[scale=0.40]{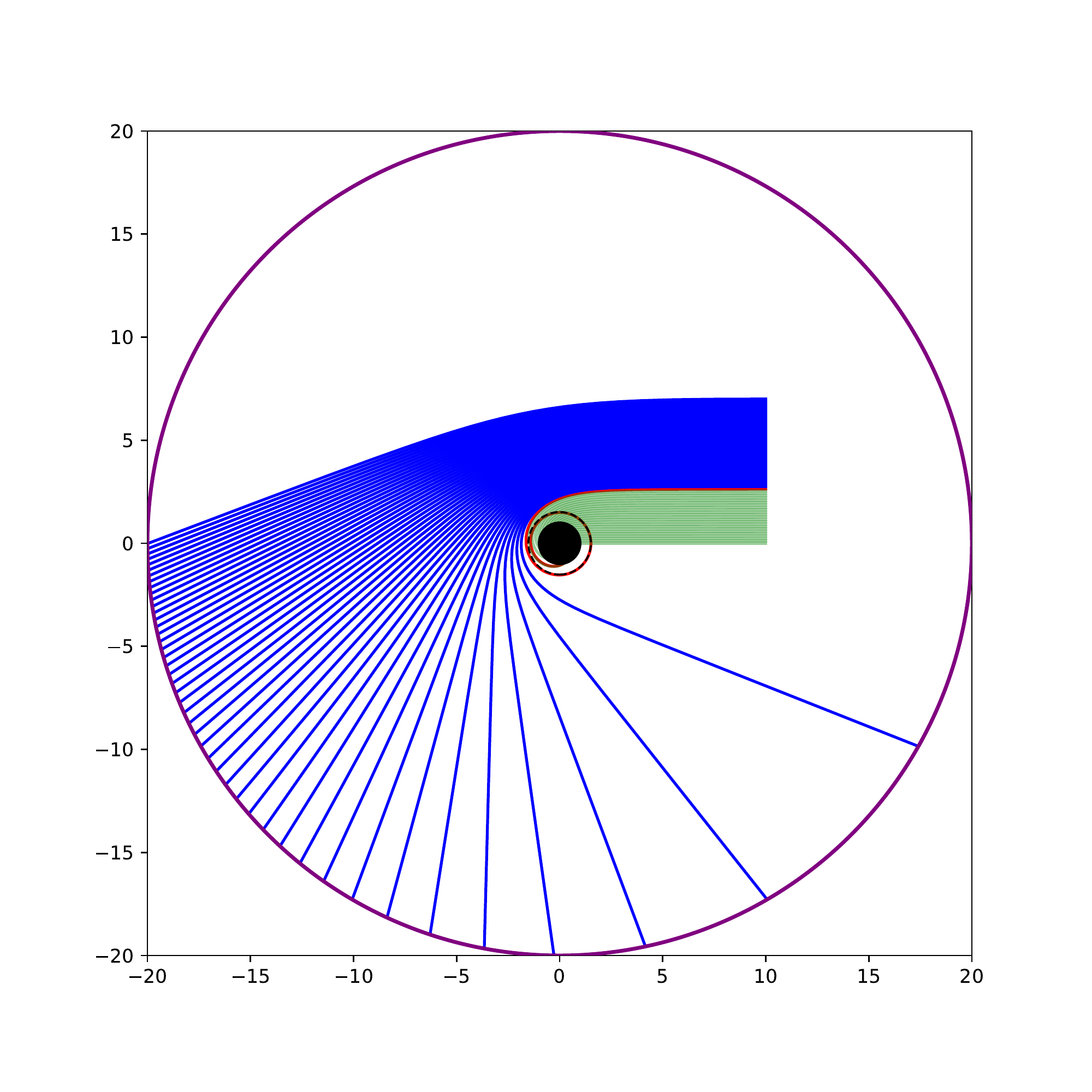}};
\node[] at (10,-40){\small  \includegraphics[scale=0.40]{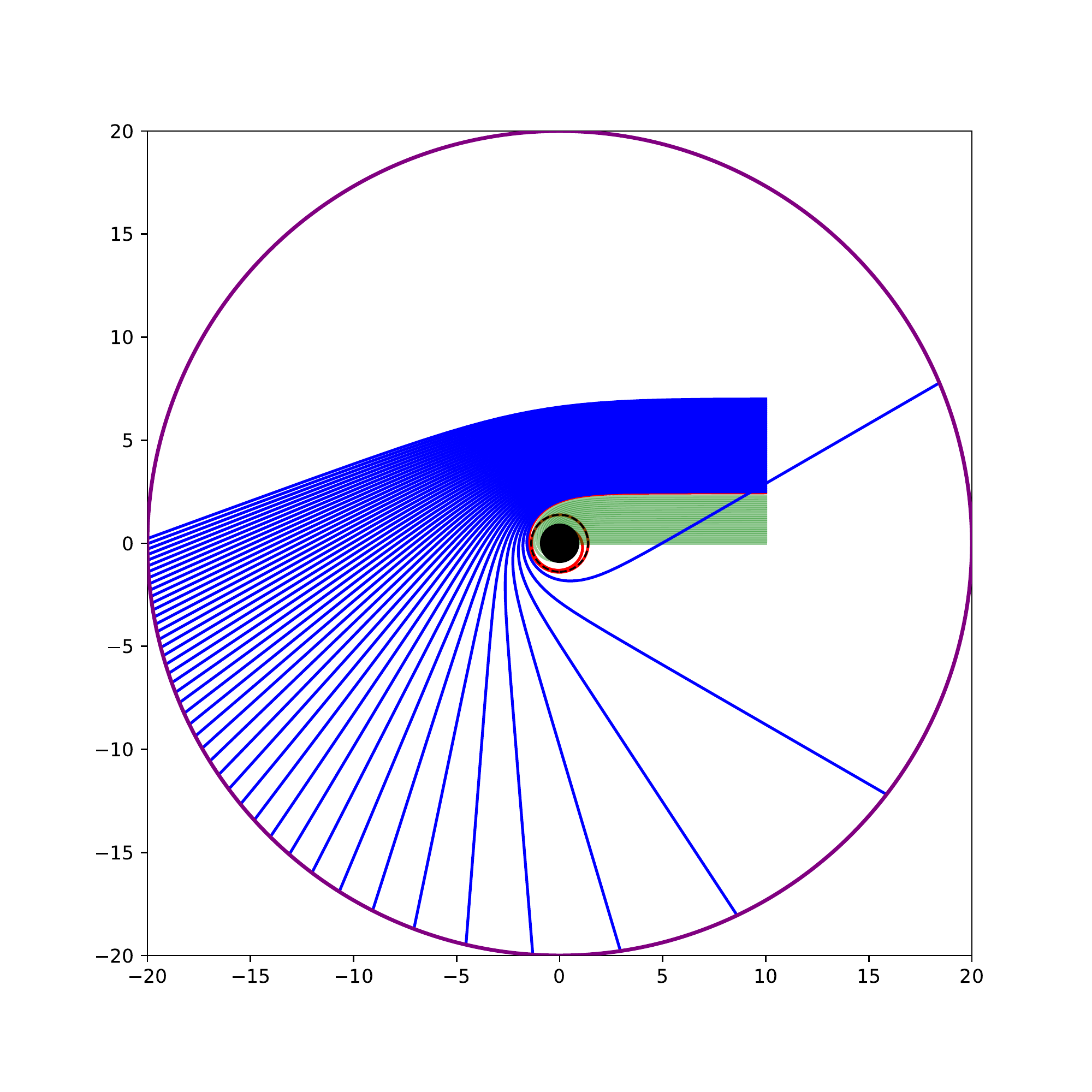}};
\node[color=black] at (-40,18.5) {$T_{cav} = 0.04\;\; Q=0.1 $};
\node[color=black] at (10,18.5) {$T_{cav} = 0.04\;\; Q=0.3 $};
\node[color=black] at (-40,-22.5) {$T_{cav} = 0.08\;\; Q=0.1 $};
\node[color=black] at (10,-22.5) {$T_{cav} = 0.08\;\; Q=0.3 $};
\end{tikzpicture}	
\caption{{\it \footnotesize The trajectories of the light ray for different values of the cavity temperature.  The horizon $r_+$ and sphere photon $r_{sp}$   are  represented by  the black disc and black dashed circle , respectively. The observer is situated in  the equatorial plane and  positioned in   $r_{ob}=15$, we take the  cavity radius  $r_{cav}=20$.}}
\label{aa}
\end{center}
\end{figure*}
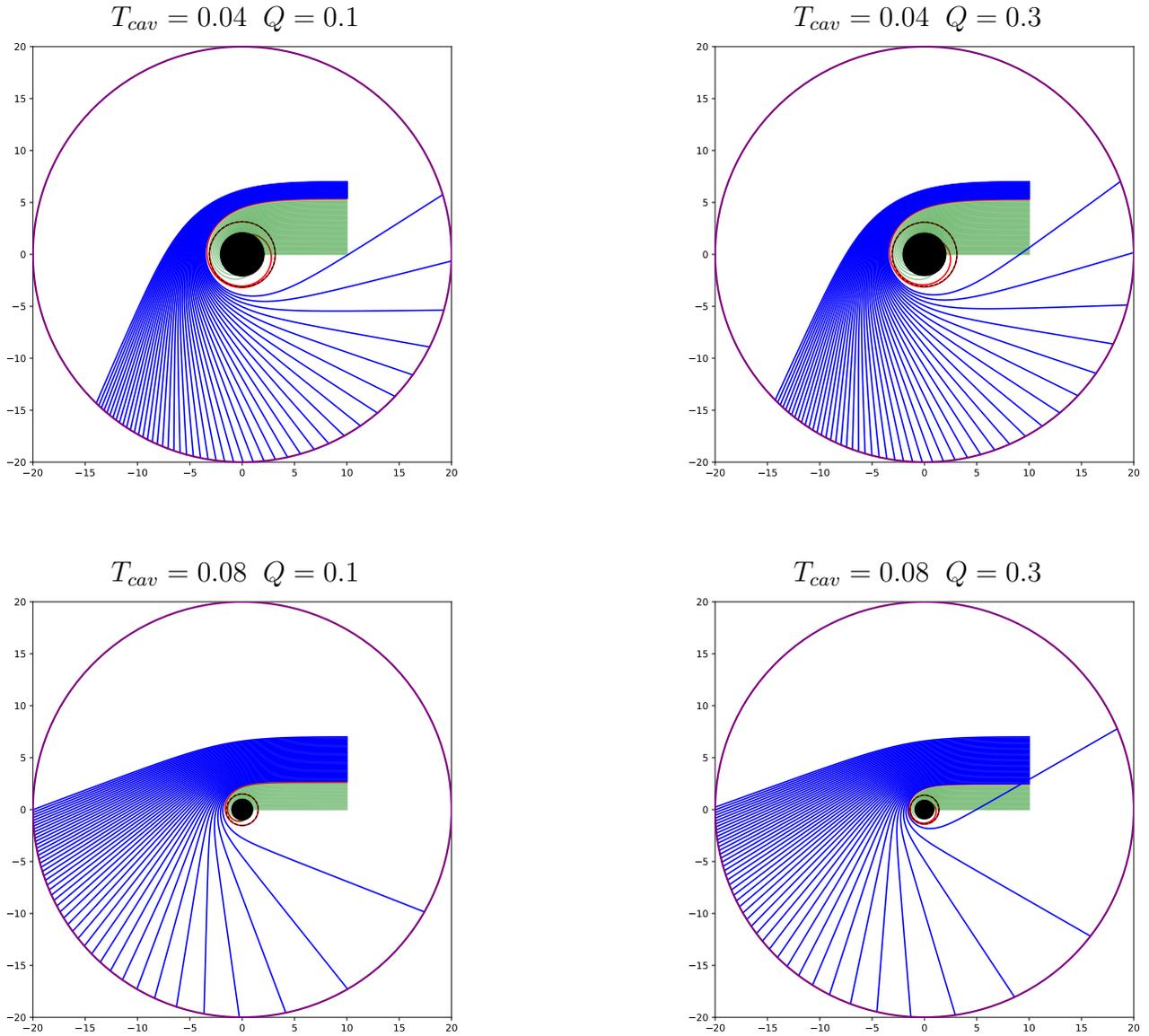
A close examination  reveals  that   the horizon and the  sphere photon radius decrease  by increasing the cavity temperature.  For fixed  values of the charge, this confirmes the  previous results associated with  the shadow and the  photon sphere  as functions of the cavity temperature.  For  $T_{cav}=0.04$,  it has been remarked that   the region  \let\textcircled=\pgftextcircled\textcircled{3} is small  than  the  region  \let\textcircled=\pgftextcircled\textcircled{1}.  An inverse behavior is observed  for  $T_{cav}=0.08$. For  such a value,   the region \let\textcircled=\pgftextcircled\textcircled{3} is large compared to the region   \let\textcircled=\pgftextcircled\textcircled{1}.   Moreover,   the region of the reflected    light rays     enlarges  with the  cavity temperature. This shows that such a temperature  can be considered as a relevant quantity modifying    the  light ray behaviors  near a black hole.  However, the charge does not affect  such trajectories. An examination shows that the light ray behaviors of  the black holes in a cavity system are different  than  the  light trajectories  around the   non rotating black holes without cavities   including the (DE) contributions\cite{a9,a16}.  It has been observed that such a distinction  comes from the cavity effect on the black hole object. 

By numerically examining the optical behavior plotted in   the above figures, we collect certain optical  results for various  moduli  space regions  in Tab.(\ref{tab:truthTables}).
\begin{table}[]
\begin{center}
\scalebox{0.7}{
\begin{tabular}{|ll|c|c|c|c|}
\hline
\multicolumn{2}{|l|}{}                                            & $T=0.02$ & $T=0.03$ & $T=0.04$ & $T=0.05$  \\ \hline
\multicolumn{1}{|l|}{\multirow{3}{*}{Q=0.1}} & $b_{sp}$                &11.75          &7.44          &5.46          &4.31                \\ \cline{2-6} 
\multicolumn{1}{|l|}{}                  & $r_{sp}$                & 6.78   &4.29          &3.15                        &2.49                  \\ \cline{2-6} 
\multicolumn{1}{|l|}{}                  & $\frac{ {\bf b_{sp}}}{ {\bf r_{sp}}}$ &   {\bf1.73 }       & {\bf1.73}          & {\bf1.73}              & {\bf1.73}                \\ \hline
\multicolumn{1}{|l|}{\multirow{3}{*}{Q=0.2}} & $b_{sp}$                &11.73          &7.42          &5.43          &4.27                \\ \cline{2-6} 
\multicolumn{1}{|l|}{}                  & $r_{sp}$                & 6.77         &4.28          &3.13                     & 2.46               \\ \cline{2-6} 
\multicolumn{1}{|l|}{}                  & $\frac{ {\bf b_{sp}}}{ {\bf r_{sp}}}$ & {\bf1.73}          & {\bf1.73}          & {\bf1.74}              & {\bf 1.74}               \\ \hline
\multicolumn{1}{|l|}{\multirow{3}{*}{Q=0.3}} & $b_{sp}$                &11.71          &7.39          &5.39           &  4.22              \\ \cline{2-6} 
\multicolumn{1}{|l|}{}                  & $r_{sp}$                & 6.76        &4.26          &3.09                      &2.42              \\ \cline{2-6} 
\multicolumn{1}{|l|}{}                  & $\frac{ {\bf b_{sp}}}{ {\bf r_{sp}}}$ &   {\bf1.73}        & {\bf1.74 }         &  {\bf1.74}             & {\bf1.75 }            \\ \hline
\multicolumn{1}{|l|}{\multirow{3}{*}{Q=0.4}} & $b_{sp}$                & 11.69         &7.34          &5.32           &4.12              \\ \cline{2-6} 
\multicolumn{1}{|l|}{}                  & $r_{sp}$                &6.73         & 4.22         &3.04                      &2.34                 \\ \cline{2-6} 
\multicolumn{1}{|l|}{}                  & $\frac{ {\bf b_{sp}}}{ {\bf r_{sp}}}$ &   {\bf1.74}        &  {\bf1.74 }        &  {\bf1.75}             & {\bf1.76  }              \\ \hline
\multicolumn{1}{|l|}{\multirow{3}{*}{Q=0.5}} & $b_{sp}$                &  11.65      &7.28          & 5.22           &3.97                 \\ \cline{2-6} 
\multicolumn{1}{|l|}{}                  & $r_{sp}$                & 6.71       &4.17          &2.97                      &2.23                  \\ \cline{2-6} 
\multicolumn{1}{|l|}{}                  & $\frac{ {\bf b_{sp}}}{ {\bf r_{sp}}}$ &   {\bf1.74}       & {\bf1.74}          & {\bf1.76 }             & {\bf1.78 }                \\ \hline

\end{tabular}
}
\end{center}
\caption{{\it \footnotesize The values of the ratio $\frac{b_{sp}}{r_{sp}}$ for different values of the cavity temperature and the charge $Q$. The observer in the equatorial plane and positioned in $r_{ob} = 15$, where one has used the cavity radius $r_{cav} = 20$.}}
    \label{tab:truthTables}
\end{table}

 According to the   reported data, we explore certain universal  optical  ratios. This could be understood as  universal optical behaviors completing the ones obtained in  the  black hole thermodynamics\cite{a39,a40}.  For small values of the charge $Q$  and the cavity temperature $T_{cav}$,  we observe  the following universal  optical ratio
\begin{equation}
\frac{b_{sp}}{r_{sp}} \sim \sqrt 3.
\end{equation} 
This universal  ratio has  not been  observed only in the black hole in a cavity  but  also in the ordinary Schwarzschild (SH) black holes without a  cavity. After a consistence  verification, we find 
\begin{equation}
\frac{b_{sp}(SH)}{r_{sp}(SH)} =\frac{3\sqrt{3}m}{3m}= \sqrt 3.
\end{equation} 
In the charged solution,  this ratio is checked to be close  to $\sqrt 3$.
\section{Conclusion and open questions}

In this work,  we have explored the optical behaviors of  the black holes in a cavity.  Concretely,  we  have  studied  the  shadows and   the photon rings  of  the   black holes surrounded by  a cavity,
where the effect of the temperature and the charge have been  examined.   After studying the effect of the horizon radius on  the observational images  of 
 the black holes by varying the charge, we have  investigated  thermal shadow behaviors   in the cavity system with certain geometric conditions.  
 For fixed charge values, we have  revealed     that the shadow radius $r_s$  increases by decreasing  the cavity temperature $T_{cav}$.  Moreover,    we have shown that  the   $r_s-T_{cav}$  curves   involve  swallowtail behaviors   observed in the $G-T$ plane of the ordinary AdS black holes. Among others,  we have found that the   $T_{HP}$ temperature in   the $G-T$   curves    coincides with the    $T_{HP}$ temperature in the $r_s-T_{cav}$ plane. The present findings  could   support the relation between the   thermodynamics    and the optical aspect of  the black holes in a cavity.

  In this way, certain thermodynamical  quantities  of the black holes including the phase transition temperature $T_{HP}$  could   be approached using the  shadow formalism  in cavity systems. Then,  we have established    the trajectory of the light rays casted  by  the black holes in  cavities.   Precisely,  we have observed  that the light  trajectories   of the  black holes in a cavity system  are  different  than the light ray behaviors of the ordinary  solutions. This distinction  comes from the cavity effect on the black hole object.  Finally,  we   have found an   optical  universal ratio associated  with  the photon sphere radius and the impact parameter.    For  small values of the charge  $Q$ and the cavity temperature $T_{cav}$,  we   have found  the optical ratio  $
\frac{b_{sp}}{r_{sp}} \sim \sqrt 3.$    This  universal optical ratio  has been verified    for the Schwarzschild  black hole without  a  cavity  which  equals to $\sqrt{3}$}.\\
This work comes up with certain questions.   This study could be adopted to various models  by enlarging the moduli space parameter.  Rotating black holes   in a cavity could be considered as possible extensions even the associated computations  will be complicated.    It should be interesting also  to consider  generic optical pictures by approaching  different aspects including the deflection angle of the  light rays. We hope address such questions in future works. 
\section*{Acknowledgments}
The authors would like to thank    N. Askour, H. El Moumni  and  Y. Hassouni,  for collaborations on related topics.  
 This work is partially
supported by the ICTP through AF.


\begin{thebibliography}{10}
\bibitem{a1}
K. Akiyama and al.,
\textit{First M87 Event Horizon Telescope Results. IV. Imaging the Central
  Supermassive Black Hole},
 Astrophys. J. {\bf L4} (1) (2019) 875, {\tt arXiv:1906.11241}.

\bibitem{a2}
K. Akiyama and al.,
\textit{First M87 Event Horizon Telescope Results. V. Imaging the Central
  Supermassive Black Hole},
 Astrophys. J. {\bf L5} (1) (2019) 875.
\bibitem{a3}
K. Akiyama and al.,
\textit{First M87 Event Horizon Telescope Results. VI. Imaging the Central
  Supermassive Black Hole},
 Astrophys. J. {\bf L6} (1) (2019) 875.
 \bibitem{a4}
A.~Belhaj, H.~Belmahi, M.~Benali, H.~El Moumni, M.~A.~Essebani and M.~B.~Sedra, {\it Optical shadows of rotating Bardeen-AdS black holes}, Mod. Phys. Lett. A \textbf{37} (2022)  2250032, {\tt arXiv:2202.10892}.
\bibitem{a5}
H.~Khodabakhshi, A.~Giaimo and R.~B.~Mann, {\it Einstein Quartic Gravity: Shadows, Signals, and Stability}, Phys. Rev. D \textbf{102} (2020) 044038, {\tt arXiv:2006.02237}. 
\bibitem{a6}
S.~W.~Wei, Y.~C.~Zou, Y.~X.~Liu and R.~B.~Mann, {\it Curvature radius and Kerr black hole shadow}, JCAP \textbf{08} (2019) 030, {\tt arXiv:1904.07710}.
\bibitem{a7}
S. Chandrasekhar, {\it The mathematical theory of black holes}, 
Oxford University Press, 1998.

\bibitem{a8}
\.I.~\c{C}imdiker, D.~Demir and A.~\"Ovg\"un, {\it Black hole shadow in symmergent gravity}, Phys. Dark Univ. \textbf{34} (2021) 100900, {\tt arXiv:2110.11904}. 

\bibitem{a9}
M.~Okyay and A.~\"Ovg\"un,{\it Nonlinear electrodynamics effects on the black hole shadow, deflection angle, quasinormal modes and greybody factors}, JCAP \textbf{01} (2022)  009, {\tt arXiv:2108.07766}. 

\bibitem{a10}
A. Övgün, I. Sakalli and J. Saavedra, {\it Shadow cast and Deflection angle of Kerr- Newman-Kasuya spacetime}, JCAP. 10 (2018) 041, arXiv:1807.00388.

\bibitem{a11}
A.~Grenzebach, V.~Perlick and C.~L\"ammerzahl, {\it Photon Regions and Shadows of Accelerated Black Holes}, Int. J. Mod. Phys. D \textbf{24} (2015) 1542024, {\tt arXiv:1503.03036}. 
\bibitem{A11}
A.~He, J.~Tao, Y.~Xue and L.~Zhang, {\it Shadow and Photon Sphere of Black Hole in Clouds of Strings and Quintessence}, Chin. Phys. C \textbf{46} (2022) 065102, {\tt arXiv:2109.13807}.
\bibitem{a12}
A.~Grenzebach, V.~Perlick and C.~L\"ammerzahl, {\it Photon Regions and Shadows of Kerr-Newman-NUT Black Holes with a Cosmological Constant}, Phys. Rev. D \textbf{89} (2014) 124004,  {\tt arXiv:1403.5234}.

\bibitem{a13}
H.~C.~D.~L.~Junior, P.~V.~P.~Cunha, C.~A.~R.~Herdeiro and L.~C.~B.~Crispino, {\it Shadows and lensing of black holes immersed in strong magnetic fields}, Phys. Rev. D \textbf{104} (2021) 044018, {\tt arXiv:2104.09577}.

\bibitem{a14}
H.~C.~D.~Lima  Junior, L.~C.~B.~Crispino, P.~V.~P.~Cunha and C.~A.~R.~Herdeiro, {\it Can different black holes cast the same shadow?},
Phys. Rev. D \textbf{103} (2021) 084040, {\tt arXiv:2102.07034}. 

\bibitem{a140} S. Vagnozzi, L. Visinelli, Hunting for extra dimensions in the shadow of M87$^*$,  Phys. Rev. D  \textbf{100} (2019) 024020, arXiv:1905.12421.


\bibitem{a141} A. Allahyari, M.  Khodadi, S.  Vagnozzi, D.  F. Mota, Magnetically charged black holes from non-linear electrodynamics and the Event Horizon Telescope, JCAP 2002 (2020) 003, arXiv:1912.08231.


\bibitem{a142}  M. Khodadi, A.  Allahyari, S. Vagnozzi, D.  F. Mota, Black holes with scalar hair in light of the Event Horizon Telescope, JCAP 2009 (2020) 026, arXiv:2005.05992.


\bibitem{a143}  C. Bambi, K. Freese, S.  Vagnozzi, L.  Visinelli, Testing the rotational nature of the supermassive object $M87^*$ from the circularity and size of its first image, Phys. Rev. D \textbf{100} (2019) 044057,   arXiv:1904.12983.

\bibitem{a15}
A.~Belhaj, M.~Benali, H.~E.~Moumni, M.~A.~Essebani, M.~B.~Sedra and Y.~Sekhmani, {\it Thermodynamic and Optical Behaviors of Quintessential Hayward-AdS Black Holes}, 
Inter.  Jour. of Geom. Meth in Mod. Phys, {\bf } (2022) 2250096, 

{\tt arXiv:2202.06290}.
\bibitem{a16}
A.~Belhaj, M.~Benali, A.~El Balali, H.~El Moumni and S.~E.~Ennadifi, {\it Deflection angle and shadow behaviors of quintessential black holes in arbitrary dimensions}, Class. Quant. Grav. \textbf{37} (2020)  215004. {\tt arXiv:2006.01078}. 
\bibitem{a17}
A.~Belhaj, H.~Belmahi and M.~Benali, {\it Superentropic AdS black hole shadows}, Phys. Lett. B \textbf{821} (2021) 136619, {\tt arXiv:2110.06771}. 
\bibitem{a18}
A.~Belhaj, H.~Belmahi, M.~Benali, W.~El Hadri, H.~El Moumni and E.~Torrente-Lujan, {\it Shadows of 5D black holes from string theory}, Phys. Lett. B \textbf{812} (2021) 13602, {\tt arXiv:2008.13478}.

\bibitem{a19}
A.~Belhaj, M.~Benali, A.~E.~Balali, W.~E.~Hadri and H.~El Moumni, {\it Cosmological constant effect on charged and rotating black hole shadows}, Int. J. Geom. Meth. Mod. Phys. \textbf{18} (2021) 2150188. {\it arXiv:2007.09058}.


\bibitem{a20}
A.~Belhaj, M.~Benali, A.~El Balali, W.~El Hadri, H.~El Moumni and E.~Torrente-Lujan, {\it Black hole shadows in M-theory scenarios}, Int. J. Mod. Phys. D \textbf{30} (2021) 2150026, {\tt arXiv:2008.09908}.

\bibitem{a21}
G.~A.~Marks, F.~Simovic and R.~B.~Mann, {\it Phase transitions in 4D Gauss\textendash{}Bonnet\textendash{}de Sitter black holes}, Phys. Rev. D \textbf{104} (2021) 104056, {\tt arXiv:2107.11352}. 
\bibitem{a22}
D. Kubiznak and R. B. Mann, {\it P-V criticality of charged AdS black holes}, JHEP \textbf{07} (2012) 033, {\tt arXiv:1205.0559}.
\bibitem{a23}
N. Altamirano, D. Kubiznak and R. B. Mann, {\it Reentrant phase transitions in rotating anti-de Sitter black holes}, Phys. Rev. D {\bf 88} (2013) 101502, {\tt arXiv:1306.5756}.
\bibitem{a24}
D. Kubiznak and F. Simovic, {\it Thermodynamics of horizons: de Sitter black holes and reentrant phase transitions}, Class. Quant. Grav. {\bf 33} (2016) 245001, {\tt arXiv:1507.08630}.
\bibitem{a26}
A.~Belhaj, M.~Chabab, H.~El Moumni, K.~Masmar and M.~B.~Sedra, {\it On Thermodynamics of AdS Black Holes in M-Theory}, Eur. Phys. J. C \textbf{76} (2016)  73, {\tt arXiv:1509.02196}.

\bibitem{a27}
A.~Belhaj, M.~Chabab, H.~El Moumni, K.~Masmar, M.~B.~Sedra and A.~Segui, {\it On Heat Properties of AdS Black Holes in Higher Dimensions}, JHEP \textbf{05} (2015)  149, {\tt arXiv:1503.07308}.
\bibitem{a28}
A.~Belhaj, A.~El Balali, W.~El Hadri, M.~A.~Essebani, M.~B.~Sedra and A.~Segui, {\it Kerr-AdS black hole behaviors from dark energy},
Int. J. Mod. Phys. D \textbf{29} (2020)  2050069.
\bibitem{a29}
A.~Belhaj, A.~El Balali, W.~El Hadri, H.~El Moumni and M.~B.~Sedra, {\it Dark energy effects on charged and rotating black holes}, Eur. Phys. J. Plus \textbf{134} (2019)  422,  {\tt arXiv:1912.08687}.

\bibitem{a290}  M. Zhang and M. Y. Guo,   {\it Can shadows reect phase structures of black holes?} Eur. Phys. J. C.
{\bf 80} (2020) 790.
\bibitem{a291} A. Belhaj, L. Chakhchi, H. El. Moumni,  J. Khalloufi, K. Masmar,  {\it Thermal image and phase transitions of charged
AdS black holes using shadow analysis}, Int. J. Mod. Phys. A. {\bf  35} (2020) 2050170.
\bibitem{a292}  X. C. Cai, Y. G. Miao,  {\it  Can we know about black hole thermodynamics through shadows?}, {\tt arXiv:2107.08352 [gr-qc]}.
\bibitem{a293}  C. Wang et al,   {\it Ruppeiner geometry of the RN-AdS black hole using shadow formalism}, Nucl. Phys. B  {\bf 976}  (2022) 115698.
\bibitem{a294}  S. Guo, G-R. Li, G-P. Li,  {\it Shadow thermodynamics of AdS black hole in regular spacetime},  {\tt  arXiv:2205.04957}. 
 \bibitem{a295}  A. Belhaj, H. Belmahi, M. Benali, A. Segui,  {\it Thermodynamics of AdS black holes from
deflection angle formalism}, Phys. Lett. B  {\bf  817} (2021) 136313.



\bibitem{a30}
H.~W.~Braden, J.~D.~Brown, B.~F.~Whiting and J.~W.~York, Jr.,{\it Charged black hole in a grand canonical ensemble}, Phys. Rev. D \textbf{42} (1990) 3376.
\bibitem{A30}
P.~Wang, H.~Wu and H.~Yang, {\it Thermodynamic Geometry of AdS Black Holes and Black Holes in a Cavity}, Eur. Phys. J. C \textbf{80} (2020)  216, {\tt arXiv:1910.07874}.
\bibitem{a31}
H.~El Moumni and J.~Khalloufi, {\it Nonlinear-Maxwell-Yukawa de-Sitter black hole thermodynamics in a cavity: I\ensuremath{-}Canonical ensemble}, Nucl. Phys. B \textbf{973} (2021) 115593.



\bibitem{a32}
H.~El Moumni and J.~Khalloufi, {\it Nonlinear-Maxwell-Yukawa de-Sitter black hole thermodynamics in a cavity: II - Grand canonical ensemble},
Nucl. Phys. B \textbf{977} (2022) 115731.


\bibitem{a33}
F.~Simovic and R.~B.~Mann, {\it Critical Phenomena of Charged de Sitter Black Holes in Cavities}, Class. Quant. Grav. \textbf{36} (2019) 014002, {\tt arXiv:1807.11875}.

\bibitem{a34}
W.~B.~Zhao, G.~R.~Liu and N.~Li, {\it Hawking\textendash{}Page phase transitions of the black holes in a cavity}, Eur. Phys. J. Plus \textbf{136} (2021)  981, {\tt arXiv:2012.13921}.


\bibitem{a35}
R.~Andr\'e and J.~P.~S.~Lemos, {\it Thermodynamics of $d$-dimensional Schwarzschild black holes in the canonical ensemble},
Phys. Rev. D \textbf{103} (2021) 064069, {\tt arXiv:2101.11010}. 

\bibitem{a36}
M.~Zhang and M.~Guo, {\it Can shadows reflect phase structures of black holes?},
Eur. Phys. J. C \textbf{80} (2020)  790, {\tt arXiv:1909.07033}.
\bibitem{Wei:2013kza}
S.~W.~Wei and Y.~X.~Liu,
{\it Observing the shadow of Einstein-Maxwell-Dilaton-Axion black hole},
JCAP \textbf{11} (2013) 063, {\tt
%doi:10.1088/1475-7516/2013/11/063
arXiv:1311.4251}.
\bibitem{a37}
X.~X.~Zeng, H.~Q.~Zhang and H.~Zhang, {\it Shadows and photon spheres with spherical accretions in the four-dimensional Gauss\textendash{}Bonnet black hole}, Eur. Phys. J. C \textbf{80} (2020)  872, {\tt arXiv:2004.12074}.

\bibitem{a38}
X.~X.~Zeng and H.~Q.~Zhang, {\it Influence of quintessence dark energy on the shadow of black hole}, Eur. Phys. J. C \textbf{80} (2020) 1058, {\tt arXiv:2007.06333}.
\bibitem{a39}  A. Belhaj, M. Chabab, H. El Moumni, M. B. Sedra,  {\it  On thermodynamics of AdS black
holes in arbitrary dimensions}, Chin. Phys. Lett.  \textbf{29} (2012) 100401,  {\tt arXiv:1210.4617}.
\bibitem{a40}  A. Belhaj, A. El Balali, W. El Hadri, E. Torrente-Lujan,   {\it   On Universal Constants of
AdS Black Holes from Hawking-Page Phase Transition}, Phys. Lett. B  \textbf{811} (2020)
135871,   {\tt arXiv:2010.07837}.
\end{thebibliography}
\end{document}